# Current Advances in Magnetoelectric Composites with Various Interphase Connectivity Types

Youness Hadouch[1,2,3,*], Daoud Mezzane[2,3], M'barek Amjoud[3], Hana Uršič[1,4], Abdelilah Lahmar[2], Brigita Rozic[1], Igor Lukyanchuk[2], Zdravko Kutnjak[1], Mimoun El Marssi[2].

**1** Jožef Stefan Institute, Jamova Cesta 39, 1000 Ljubljana, Slovenia.

**2** University of Picardie Jules Verne, Scientific Pole, 33 rue Saint-Leu, 80039 Amiens Cedex 1, France.

**3** Cadi-Ayyad University, Faculty of Sciences and Technology, BP 549, Marrakech, Morocco.

**4** Jožef Stefan International Postgraduate School, Jamova cesta 39, 1000 Ljubljana, Slovenia

***Corresponding author:**

E-mail: hadouch.younes@gmail.com; youness.hadouch@ijs.si

ORCID: https://orcid.org/0000-0002-8087-9494

Tel: +212-6 49 97 06 74

## Abstract:

Magnetoelectric composites integrate the coupling between magnetic and piezoelectric materials to create new functionalities for potential technological applications. This coupling is typically achieved through the exchange of magnetic, electric, or elastic energy across the interfaces between the different constituent materials. Tailoring the strength of the magnetoelectric effect is primarily accomplished by selecting suitable materials for each constituent and by optimizing geometrical and microstructural designs. Various composite architectures, such as (0-3), (2-2), (1-3) and core-shell connectivities, have been studied to enhance magnetoelectric coupling and other required physical properties in composites. This review examines the latest advancements in magnetoelectric materials, focusing on the impact of different interphase connectivity types on their properties and performance. Before exploring magnetic-electric coupling, a brief overview of the historical background of multiferroic magnetoelectric composites is provided. Fundamental concepts underlying the magnetoelectric effect, piezoelectricity, and the magnetostrictive effect are explained, including their origins and examples of these materials' properties. So far, three types of magnetoelectric composite connectivities have been investigated experimentally: particulate composites (0-3), laminated and thin films (2-2), sticks embedded in matrix, core-shell particles, and coaxial fibers. An outlook on the prospects and scientific challenges in the field of multiferroic magnetoelectric composites is given at the end of this review.

## Graphical abstract

Graphical abstract representing different connectivity in composite multiferroic materials.

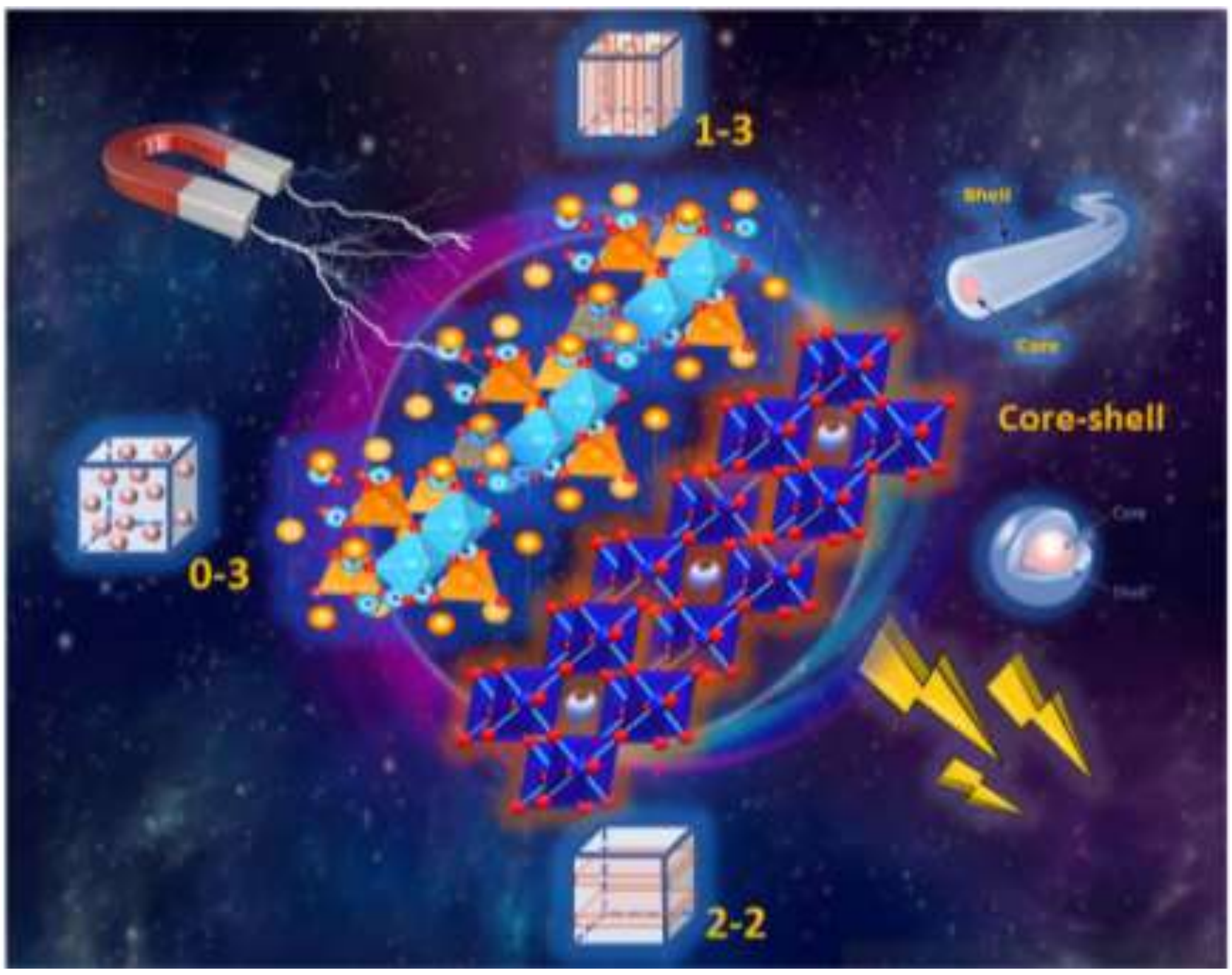

Contents

## List of Abbreviations

MF: Multiferroic

ME: Magnetoelectric

FE: Ferroelectric

P: Polarization

M: Magnetization

E: Electric field

H: Magnetic field

MeRAM: Magnetoelectric Random Access Memory

BT: $BaTiO_3$

RT: Room temperature

$T_C$: Transition temperature

B: Magnetic induction

MEVC: Magnetoelectric voltage coefficient

DME: Direct magnetoelectric effect

$P_r$: Remnant polarization

MDE: Magneto-dielectric effect

EMR: Electromechanical resonance

LGT: $La_3Ga_{5.5}Ta_{0.5}O_{14}$

CME: Converse magnetoelectric effect

FMR: Ferromagnetic resonance

PFM: Piezo-response force microscopy

MFM: magnetic force microscopy

SMM: Scanning microwave microscope

PT: Lead Titanate $PbTiO_3$

PZT: Lead Zirconate Titanium $PbZr_{1-x}Ti_xO_3$

PMN: Lead Magnesium Niobate

KNN: $K_{0.5}Na_{0.5}NbO_3$

BNT: $Bi_{0.5}Na_{0.5}TiO_3$

BFO: $BiFeO_3$

O: Orthorhombic

T: Tetragonal

R: Rhombohedral

MPB: Morphotropic phase boundary

BZT: $BaZr_{1-x}Ti_xO_3$

BCT: $Ba_xCa_{1-x}TiO_3$

BCZT: $Ba(Ti_{0.8}Zr_{0.2})O_3$–$(Ba_{0.7}Ca_{0.3})TiO_3$

$K_p$: Electromechanical coupling coefficient

$g_{33}$: Piezoelectric voltage coefficient

EMF: Electromotive force

RE: Rare earth

CFO: $CoFe_2O_4$

FeRAM: Ferroelectric Random-Access Memory

DM: Dzyaloshinskii–Moriya

NFO: $NiFe_2O_4$

NZFO: $Ni_{1-x}Zn_xFe_2O_4$

LSMO: $La_{1-x}Sr_xMnO_3$

LCMO : $La_{1-x}Ca_xMnO_3$

KNNS–BNKH : $(1-x)(K_{1-y}Na_y)(Nb_{1-z}Sb_z)O_3$–$xBi_{0.5}(Na_{1-w}K_w)_{0.5}HfO_3$

Terfenol-D: $Tb_{1-x}Dy_xFe_2$

BST: $Ba_{0.9}Sr_{0.1}TiO_3$

PVDF: Polyvinylidene fluoride

P(VDF-TrFE): Polyvinylidene fluoride-trifluoroethylene

P(VDF-HFP): Polyvinylidene fluoride-hexafluoropropylene

BTS: $BaTi_{1-x}Sn_xO_3$

BCTSn: $Ba_{0.95}Ca_{0.05}Ti_{0.89}Sn_{0.11}O_3$

AAO: Anodic Aluminium Oxide

MENPs: Magnetoelectric nanoparticles

MRI: Magnetic resonance imaging

# I Introduction

Recently, the microelectronics industry has developed increasingly small integrated circuits with even more sophisticated functions with the tendency toward miniaturizing devices to increase speed, reduce power consumption, and lower cost [1]. This trend promotes the development of multifunctional materials, which combine the well-known characteristics into a single component [2]. Among them, multiferroic materials (MF) with coexistence of at least two ferroic orders, namely ferroelectric (FE) with spontaneous polarization, ferro(antiferro/ferri) magnetic with spontaneous magnetization and ferroelastic with a spontaneous deformation, have attracted more attention as a result of their potential for applications as multifunctional devices [3,4]. Interesting outcomes, like the magnetoelectric (ME) effect, can result from coupling between these ferroic orders as schematically represented in Figure 1 [5,6]. Therefore, ME materials have the ability to switch and/or tune the polarization (P) by a magnetic field (H) and to control the magnetization (M) by an electric field (E) [7]. This property provides a new perspective on the upcoming generation of innovative electronic devices [8,9].

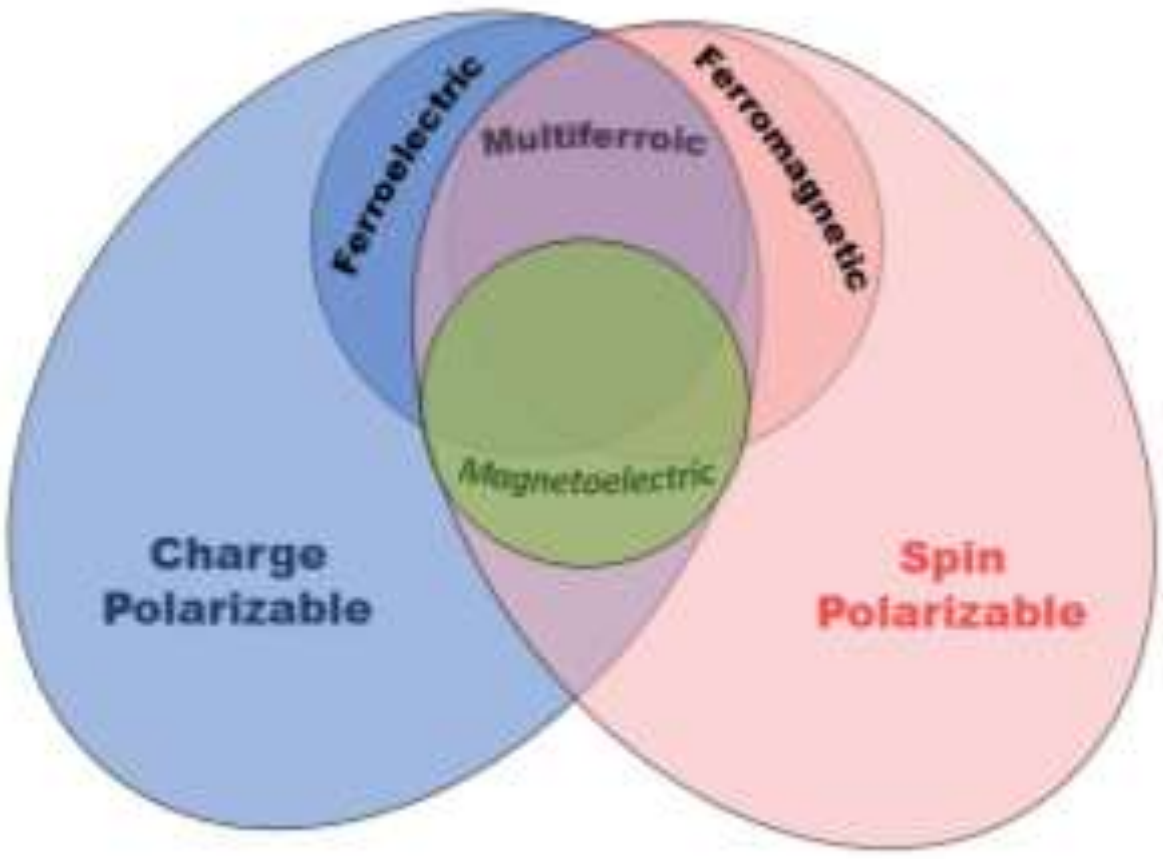

***Fig. 1.*** *Schematic diagram for a MF and magnetoelectric ordering in materials driven/controlled by ferroelectric and ferromagnetic behavior of the materials (After [10]).*

Magnetoelectric materials have been actively investigated over the last few decades because of their many potential applications [9,11,12]. The coexistence of several order parameters and ME coupling can be exploited simultaneously to control or manipulate these properties with an electric or magnetic field. One of the most promising applications of these materials in electronics is the encoding of magnetic information into an MeRAM memory (Magnetoelectric Random Access Memory) using an electric field. With this approach, by utilizing ME coupling, we may write four polarization states rather than two resulting in a considerable energy gain over simply producing a magnetic field [13,14]. However, MeRAM is still in theoretical studies and requires further research to be realized [15]. In addition, this approach will have the potential to increase the amount of storable data without reducing storage surfaces [13,15]. Furthermore, by combining electrical and magnetic properties in a monolithic material, the induced multiferroicity has the potential to offer cutting-edge applications, including enhancing computing power in electronic devices, reducing energy consumption, and minimizing waste. This is why, multiferroic materials are used in the field of sensors [16–18], memories [13,19–21], current/voltage convertors [22–24], energy-related applications, such as energy harvesters [25–27], energy storage devices [28–31], and cooling applications based on caloric effects [32–35] as well as electric field tunable devices, and could be used in other multifunctional devices. In biology and medicine, ME materials are also used in brain stimulation [36,37], medication delivery [38,39], tissue engineering [40,41], and wireless power transfer [42,43]. Some other applications that are still in the prototype stage include hydrogen sensor detection, geomagnetic field sensors, tunable microwave devices, and gyrators [44–47]. Figure 2 depicts some key applications of ME composites.

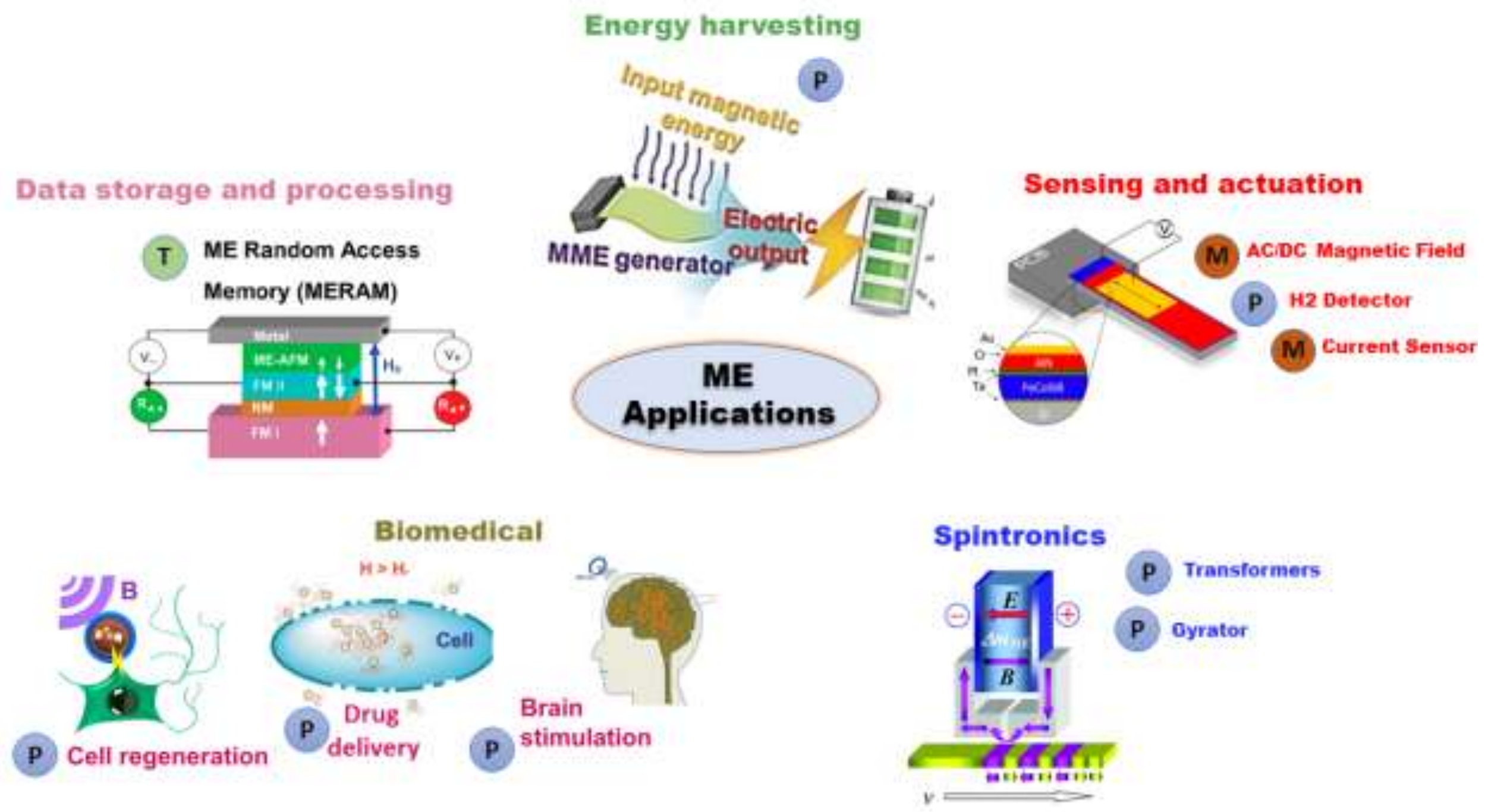

***Fig. 2.*** *Main applications of ME materials, the letters beside each application mean [M: the device is in the market, P still prototype and T means still theoretical study] (After* [48–51]*).*

In 1894, P. Curie started the first historical discussion regarding the existence of a linear relationship between the electrical and magnetic properties, i.e., that some natural materials can be both magnetized and polarized by a magnetic field and an electric field [52].

Based on symmetry considerations, Landau and Lifshitz predicted the possible presence of magnetoelectric coupling in specific crystals [53]. After several studies on various perovskites, researchers identified nickel iodine boracite ($Ni_3B_7O_{13}I$) as the first compound in which ferroelectricity and ferromagnetism coexist [54]. However, despite this observation, cross-coupling interaction between ferroelectricity and ferromagnetism was not achieved in this material. The study of magnetoelectric materials gained popularity in the 1960s thanks to the discovery and exploitation of the properties of ferroelectric materials like $Cr_2O_3$ [55] and $BaTiO_3$ (BT) [56] in the 1940s, following Dzyaloshinskii's generic prediction one year earlier [57]. In 1974, particulate composites containing either ferroelectric and magnetic particles were extensively studied [58]. However, the low magnetoelectric coupling observed in these composites as well as the high dielectric losses, have limited their practical applications. A new wave of interest in magnetoelectricity emerged in the 2000s, along with new advanced experimental methods for the synthesis and characterization of nanomaterials. Starting currently, new multiferroic composites exhibiting various connectivities [e.g., laminated composites, vertically aligned nanocomposites and nanoporous materials] have been emerged, including organic compounds with piezoelectric properties [58,59]. Figure 3 illustrates how the number of scientific publications increased from 601 to 17500 between 2000 and 2024, according to Google Scholar.

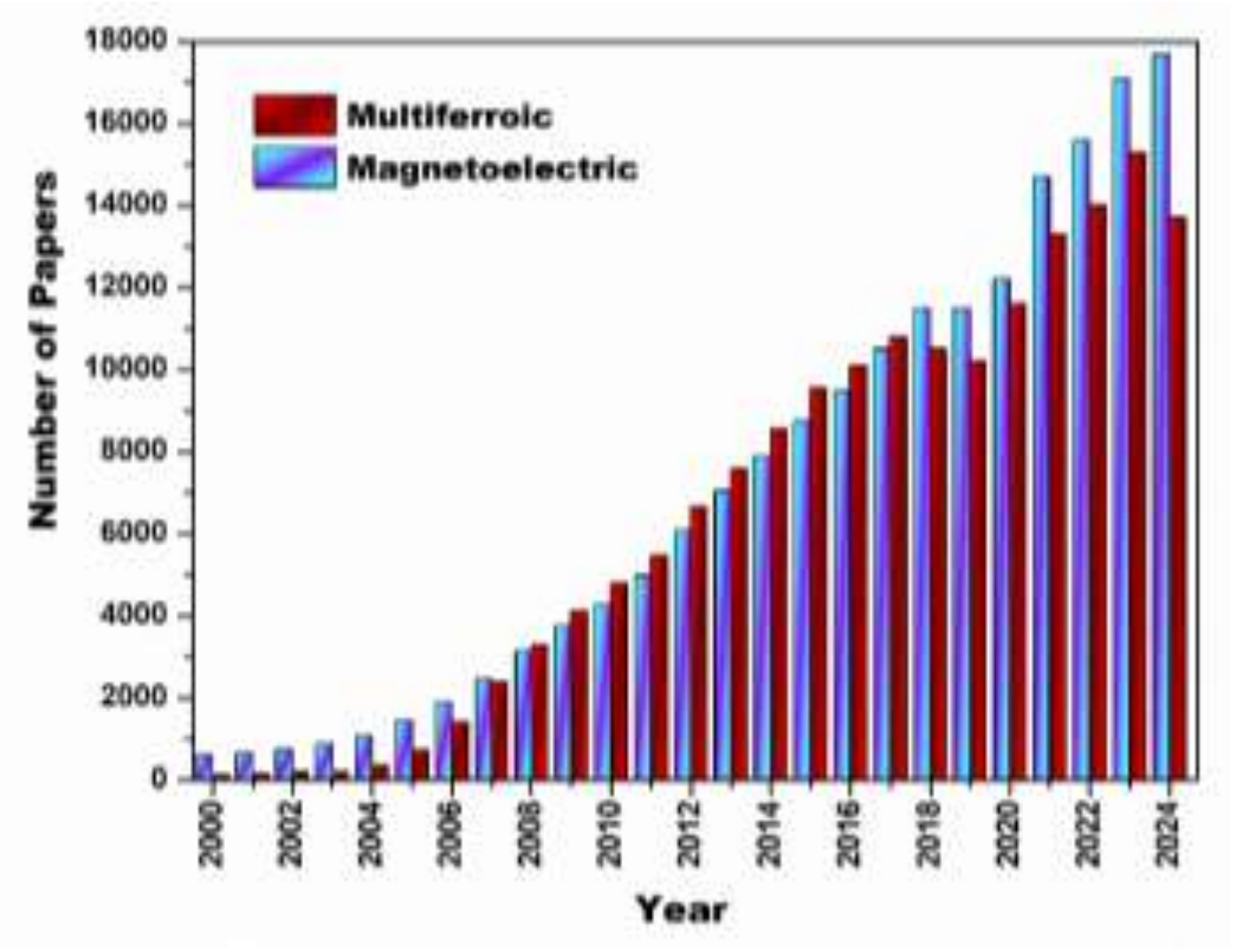

***Fig. 3.*** *The total number of articles reported in Google Scholar from 2000 to 2024 pertaining "multiferroic" and "magnetoelectric".*

Monolithic magnetoelectric materials, where magnetic and electric fields are intrinsically coupled, are characterized by their chemical homogeneity and isotropy. The coexistence of magnetic and polar sublattices in this type of ME material is intriguing from a fundamental standpoint [59]. However, the challenge arises from the inherent mutual exclusion of ferromagnetism and ferroelectricity, making it difficult to find single materials with ME coupling at room temperature (RT).

To date, only a limited number of single-phase materials demonstrating measurable coupling at room temperature have been identified, including $BiFeO_3$ [60–63], $BiMnO_3$ [64], $YMnO_3$[65], $TbMnO_3$ [66,67], $Pb(Fe_{0.5}Nb_{0.5})O_3$ [68], and $Fe_3O_4$ [69]. However, the majority of these phases exhibit a magnetic transition temperature ($T_C$) at very low temperature, which differs significantly from the FE transition temperature. In addition, they show low permittivity or low permeability at room temperature resulting in a minimal ME coupling.

Nevertheless, the single-phase ME materials are unsuitable for envisioned devices due to their low ME coupling coefficient and lower operational temperature. To overcome these limitations, single-phase materials with appropriate dopants have been involved in the A-site, B-site, and/or both perovskite sites. Nonetheless, despite some improvements in ME coupling, their practicality for device applications is still restricted. The coupling that has been identified in this class of single-phase materials varies from 1 to 20 mV $cm^{-1}$ $Oe^{-1}$ [70].

To achieve high ME coupling with transition temperatures (ferroelectric and magnetic) above RT, researchers have explored artificial ME composite structures. These composite structures involve ferroelectric materials characterized by substantial polarization, high piezoelectric coefficients, elevated $T_C$ (well above RT), and magnetic materials with significant magnetization, high resistivity, and substantial magnetostrictive coefficients. Such combinations could result in substantial ME coupling at room temperature [71–76]. Multi-phase composites feature spatially distinct piezoelectric and magnetostrictive phases connected via an interface. Within these composites, magnetoelectric coupling is indirectly induced by the strain interaction between the piezoelectric and magnetostrictive effects [49].

There is a large variety of geometry available for combining the magnetic and electric phases in a multiferroic composite since the two phases of the composites are spatially separated. It refers to this as connectivity [77,78]. The concept of phase connectivity, introduced by Newnham [79], describes the structure of two-phase composites using notations like 0-3 called particulate composite, 2-2 are films, laminate ceramic or horizontal heterostructures, and vertical heterostructures (fibers or rods) denoted by 1-3, etc., where each number represents the connectivity of each phase. For instance, a particulate composite designated as 0-3 comprises single-phase nanoparticles (indicated by 0) embedded within a matrix of a distinct phase (indicated by 3). Figure 4 depicts the diagrams of the four major types of composite connectivity types.

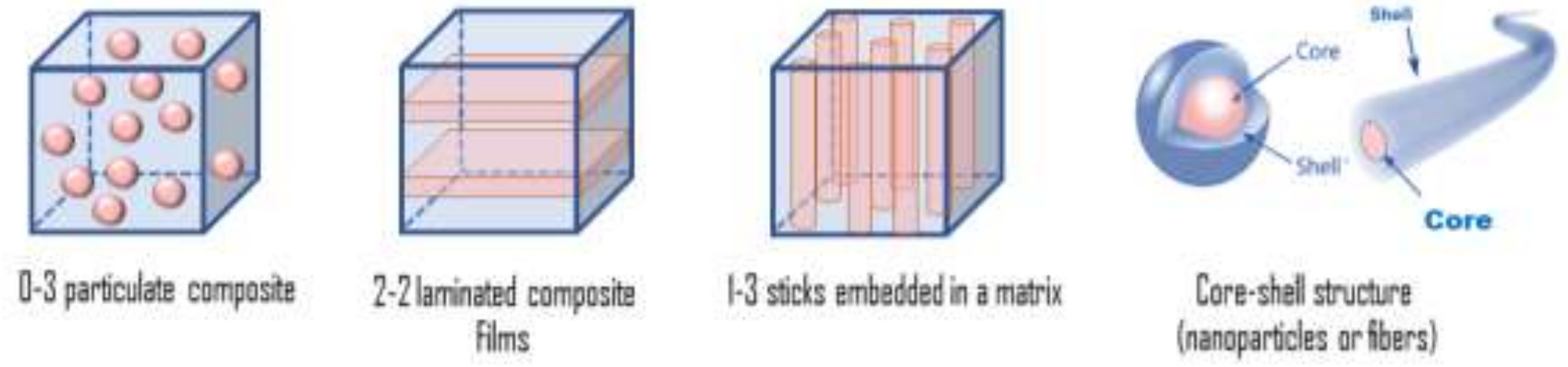

***Fig. 4.*** *Diagrams of the four major types of composite connectivity arrangements.*

By carefully selecting and engineering the connectivity between the ferroelectric and magnetic components, the mechanical, electrical, and magnetic signal transfer inside the composite can be significantly improved. In fact, well-designed connectivities enable the synergistic interaction between the ferroelectric and magnetic materials, leading to enhanced performance characteristics such as increased ME coupling strength and improved stability of the composite structure, with the ability to tune and control their magnetoelectric properties for specific external stimuli and applications. Furthermore, the interconnection of polarization topological excitations emerging in the embedded nanopieces may substantially influence the functional properties of the composites [80].

## II Fundamentals of Strain-Mediated Magnetoelectric Effect

### 1. Magnetoelectric effect

The discovery of Maxwell's equations in 1865 challenged the initial belief that magnetism and electricity were independent phenomena, suggesting instead that they were closely related [81]. A linear relationship between these two orders is discussed by Pierre Curie in 1894, proposing that some materials in nature can be polarized by a magnetic field and can be magnetized by an electric field [52]. According to the Landau free energy formulation, the ME effect might be the linear or nonlinear coupling between the electrical and magnetic order parameters [82] (eq. 1):

$$\mathrm{G(E,H)} = G_0 - P_i^S E_i - M_i^S H_i - \frac{1}{2}\varepsilon_0\varepsilon_{ij}E_iE_j - \frac{1}{2}\mu_0\mu_{ij}H_iH_j - \alpha_{ij}E_iH_j - \frac{1}{2}\beta_{ijk}E_iH_jH_k - \frac{1}{2}\gamma_{ijk}H_iE_jE_k - \frac{1}{2}\delta_{ijkl}E_iE_jH_kH_l - \cdots$$

Where G is Gibbs free energy, (i, j, k) refer to the three components of variables in spatial coordinates, E is electric field, H is magnetic field, $P^S$ is spontaneous polarization, $M^S$ is spontaneous magnetization, ε is electric susceptibility, μ is magnetic susceptibility, α is linear ME coefficient, and β, γ, and δ are higher order of ME coefficients.

The ME effect observed in composite systems results from the interaction between the electrical and magnetic order parameters of FE and magnetic phases. While individual ferroelectric and magnetic phases do not display the ME effect, the combined hybrid composite system comprising both phases demonstrates significant ME coupling [83]. It is noteworthy that the ME coupling can be either direct or converse (indirect), depending on the elastic interactions and nature of the applied field (electric or magnetic).

### 2. ME Coupling Coefficient

The ME coupling can be classified into two types: direct and converse ME coupling, depending on the elastic interactions and applied electric or magnetic field [84].

$$Direct\ ME\ coupling = \frac{\text{magnetic}}{\text{mechanical}} \times \frac{\text{mechanical}}{\text{electrical}}$$

$$Converse\ ME\ coupling\ = \frac{\text{electrical}}{\text{mechanical}} \times \frac{\text{mechanical}}{\text{magnetic}}$$

In direct ME coupling, the coupling takes place via mechanical strain transmission between the FE and magnetic phases, while in converse ME coupling, the coupling takes place via an elastic interaction through the inverse electrostrictive/piezoelectric and magneto-strictive/piezomagnetic effect between the FE and magnetic phases [84]. Figure 5 illustrates the strain-mediated ME effect in multiferroic composites.

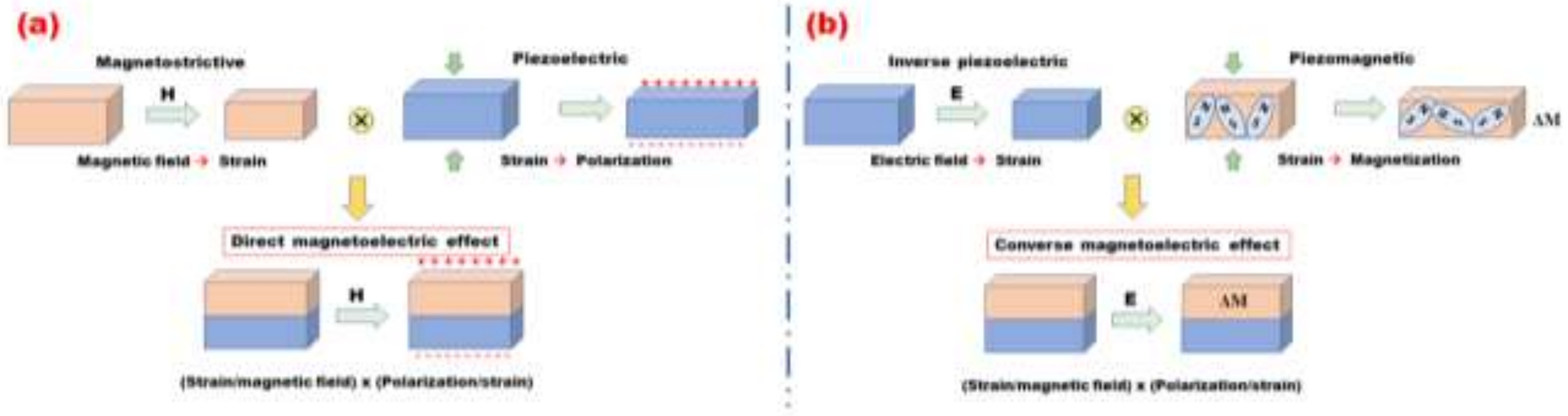

***Fig. 5.*** *Schematic illustration of strain-mediated ME effect in a composite system consisting of a magnetic layer (beige) and ferroelectric layer (blue). (a) Direct ME effect and (b) converse ME effect (After [85]).*

Several mechanisms are involved in converse ME coupling, such as strain, variation of spin-polarized charge densities, modulation of interfacial oxidation, and coupling due to spin exchange. The strain-mediated ME coupling is the most common mechanism, where the strain generated by the magnetic field induces a change in the polarization of the ferroelectric phase. However, generating and transmitting strain in composite structures is not always easy to achieve. The ME coupling coefficient can be enhanced by optimizing the connectivity, conductivity, poling, and bias field of the composites [86].

The magnetoelectric effect (ME) can be represented by ME coupling coefficient (α). When an electric field E is applied, the material's change in magnetic induction B is described by electrically induced ME coupling [87].

$$\alpha_{ij}^{E} = \frac{\partial B_i}{\partial E_j} \quad \text{(eq. 2)}$$

By introducing a magnetic field (H), magnetic coupling defines the change in electrical polarization (P):

$$\alpha_{ij}^{H} = \frac{\partial P_i}{\partial H_j} \quad \text{(eq. 3)}$$

E = (V/e); where V is the voltage and e is the thickness of the sample.

Thus, the magnetically produced ME effect can be described as follows:

$$\alpha_{ij}^{H} = \left(\frac{\partial P_i}{\partial H_j}\right) = \varepsilon_0 \varepsilon_{ii} \left(\frac{\partial E_i}{\partial H_j}\right) = \frac{\varepsilon_0 \varepsilon_r}{e} \left(\frac{\partial V}{\partial H}\right) = \varepsilon_0 \varepsilon_r \alpha_V^H \quad \text{(eq. 4)}$$

Where $\alpha_V^H$ is the magnetically induced ME voltage coefficient (MEVC)

$$\alpha_V^H = \left(\frac{\partial E}{\partial H}\right) = \frac{1}{e}\left(\frac{\partial V}{\partial H}\right) \text{ (eq. 5)}$$

The MEVC is the most commonly used parameter for analyzing experimental data [88].

The relation between ME coupling coefficient ($\alpha^H$) and MEVC ($\alpha_V^H$) is as follows:

$$\alpha^H = \varepsilon_0 \varepsilon_r \alpha_V^H \quad \text{(eq. 6)}$$

Both $\alpha^H$ and $\alpha^E$ are expressed as [s m$^{-1}$] in SI unit, whereas MEVC is expressed in [V A$^{-1}$] in SI units and [V cm$^{-1}$ Oe$^{-1}$] in CGS units [87,88].

$$\alpha^H \propto d.q \quad \text{(eq. 7)}$$

Where *d and q* represent the piezoelectric and piezomagnetic coefficients, respectively. *q=dλ/dH* (*λ* denotes the magnetostriction of the ME material).

## 3. Measurements of ME effect

### a. Measurement of direct magnetoelectric effect DME

The strength of DME could be quantified using three methods:

-By calculating the MEVC at low frequencies and at frequencies corresponding to mechanical resonance in the ME composites [89];

- By determining the change in the remnant polarization ($P_r$) calculated from P versus E data under magnetic field H [90];

-By calculating the magneto-dielectric effect (MDE) [91].

The most typical method for measuring MEVC is to apply an AC magnetic field (δH) and measure the induced voltage (δV) in the composite. The MEVC is determined using equations discussed above. Preceding measurement, the ferroelectric phase of the composite should be poled by applying an electric field. Then, an Ac magnetitic field is applied in addition to an DC bias H so that the MEVC is maximized, as shown in Figure 6a. When H=0, the AC magnetostriction λ (mechanical deformation) remains small and $\alpha_V^H$ ~ 0. However, with the introduction of a bias field, the AC magnetostriction can be significantly enhanced. Furthermore, Figure 6a shows that as magnetostriction reaches saturation $\alpha_V^H$ approaches zero.

The MEVC ($\alpha_V^H$) is then measured as a function of H, and the variation in δE with H corresponds to the change in q (slope of λ vs H). While H and δH can be applied at different orientations relative to the sample geometry, the MEVC achieves its maximum when the orientation corresponds to minimum demagnetization [92].

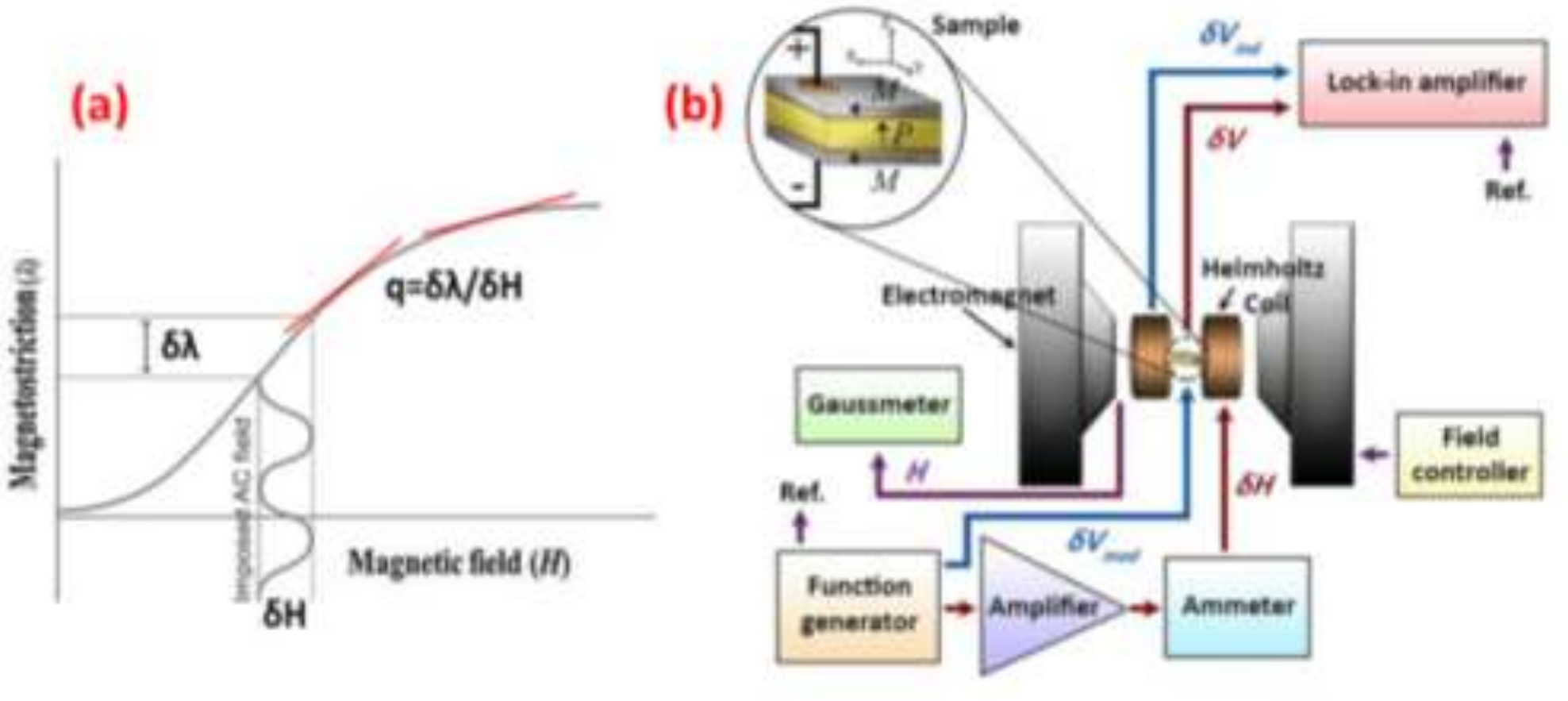

***Fig. 6.*** *(a) Schematic diagram depicting typical magnetostriction versus bias field H for a ferromagnet. For low-frequency measurements of the MEVC under an AC magnetic field dH, the bias field H is necessary to achieve a maximum of the piezomagnetic coefficient q (after [93]), (b) Schematic of the experimental setup for measuring direct magnetoelectric effect [94].*

Figure 6b illustrates a set up employed to measure MEVC. The setup involves subjecting the sample to a bias field H generated by an electromagnet, as well as an AC field δH produced by a pair of Helmholtz coils. Notably, the coils should not be wound or mounted on the pole pieces of the magnet. To prevent the sample from picking up noise, it must be shielded within a metal box. Measurement of the differential voltage across the sample necessitates the use of a three-terminal network. For precise determination of MEVC, employing lock-in detection is recommended. MEVC is observed relative to variations in the bias field H, frequency and amplitude of δH, applied field orientations, and temperatures [94].

A phenomenon of both fundamental and technological significance in the field of ME materials is the coupling observed when the composite exhibits resonant behavior, such as bending resonance or longitudinal/thickness electromechanical resonance (EMR) [95]. This resonance ME effect shares similarities with the conventional effect, where an induced polarization occurs in response to an alternating magnetic field. However, in this case, the alternating field is precisely adjusted to the acoustic frequency, resulting in a substantial increase in the MEVC. For example, to measure the ME response during mechanical resonance, the Metglas (FeBSiC), and piezoelectric langatate, $La_3Ga_{5.5}Ta_{0.5}O_{14}$ (LGT) sample can be stimulated with a pulsed magnetic field, as illustrated in Figure 7a, and the generated voltage versus frequency spectra will display peaks corresponding to the resonance modes, as depicted in Figure 7b. Additionally, the determination of the DME effect could be also done by measuring polarization versus electric field data under a static magnetic field. The change in $P_r$ induced by the magnetic field serves as a measure of the strength of ME interactions (Fig. 7c). Furthermore, the magneto-dielectric effect, which involves observing changes in permittivity (composite capacitance) with a static magnetic field, is utilized for investigating ME coupling in composites (Fig. 7d) [90,96–98].

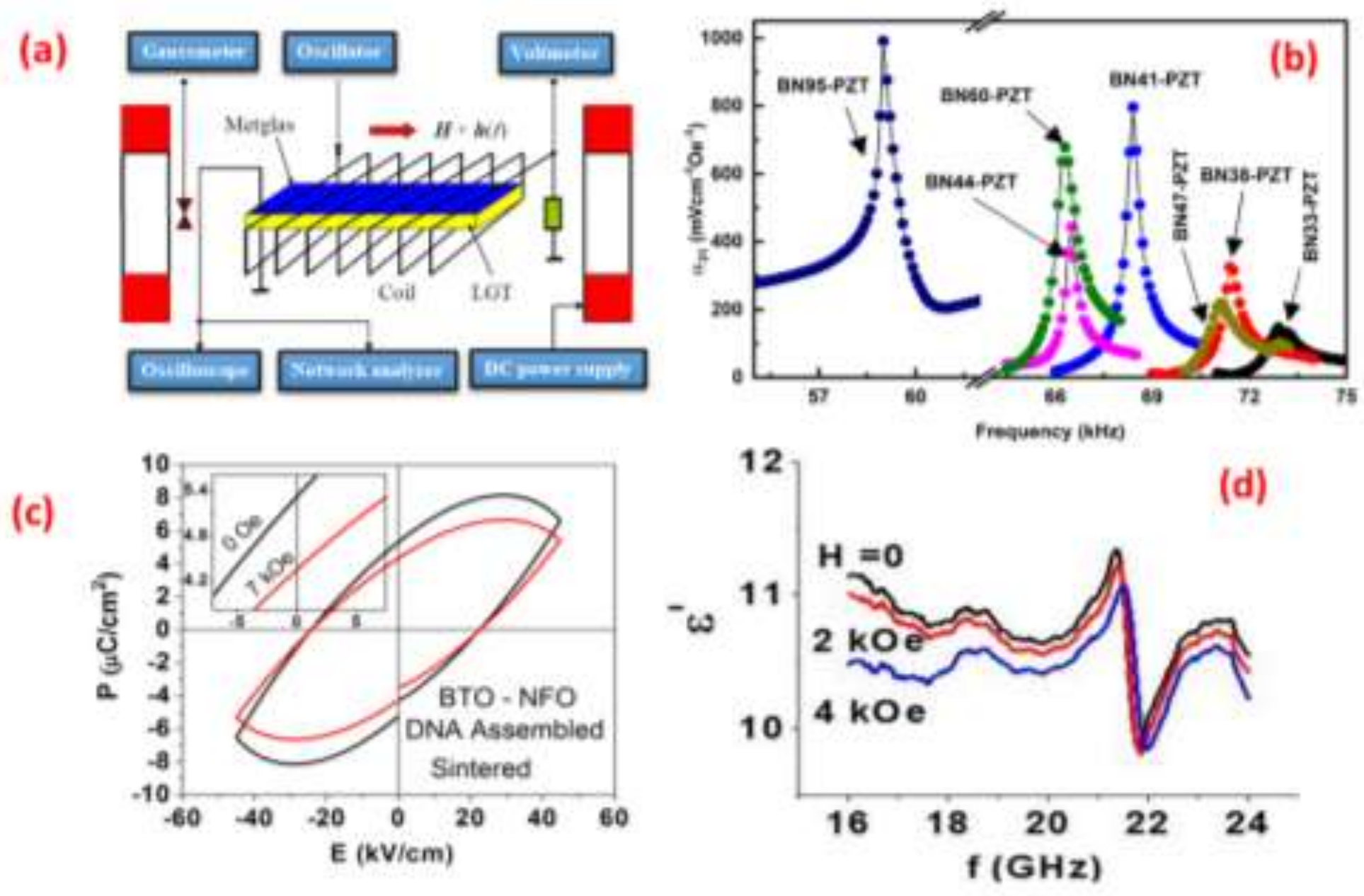

***Fig. 7.*** *(a) Experimental set up to measure frequency f dependence of MEVC by passing a pulsed current through a coil and obtaining frequency spectra of the ME voltage (after [97]). (b) Frequency dependence of ME coefficient $\alpha_{31}$ for bilayer of BNx-PZT. The peak values of MEVC occur at longitudinal mechanical resonance frequency in the samples [98],(c) RT P-E hysteresis loops without and with an external magnetic field [90] (d) dependence of relative dielectric permittivity on frequency under magnetic field [96].*

### b. Measurement of converse magnetoelectric effect CME

The CME effect is produced by applying an external AC electric field to a magnetoelectric composite, causing the piezoelectric phase to distort and change the composite's magnetic properties. Small magnetic fields may be detected by winding a pick-up coil around the composite and measuring the induced voltage in the coil as it oscillates at its mechanical resonant frequency. By analogy, the strength of the ME coupling is measured in terms of $\alpha_B^E$= B/E. Hayes et al. reported that this approach could produce field sensitivity of up to 64 kV $T^{-1}$ [99].

A simple method to calculate the ME coupling is by determining the slope of the linear graph of induced magnetization versus AC voltage amplitude [88]. Alternative methods frequently utilized for CME include ferromagnetic resonance (FMR) or measurements of magnetization M versus applied magnetic field (M vs H using a vibrating sample magnetometer) while an electric field is applied as shown in Figure 8a [100].

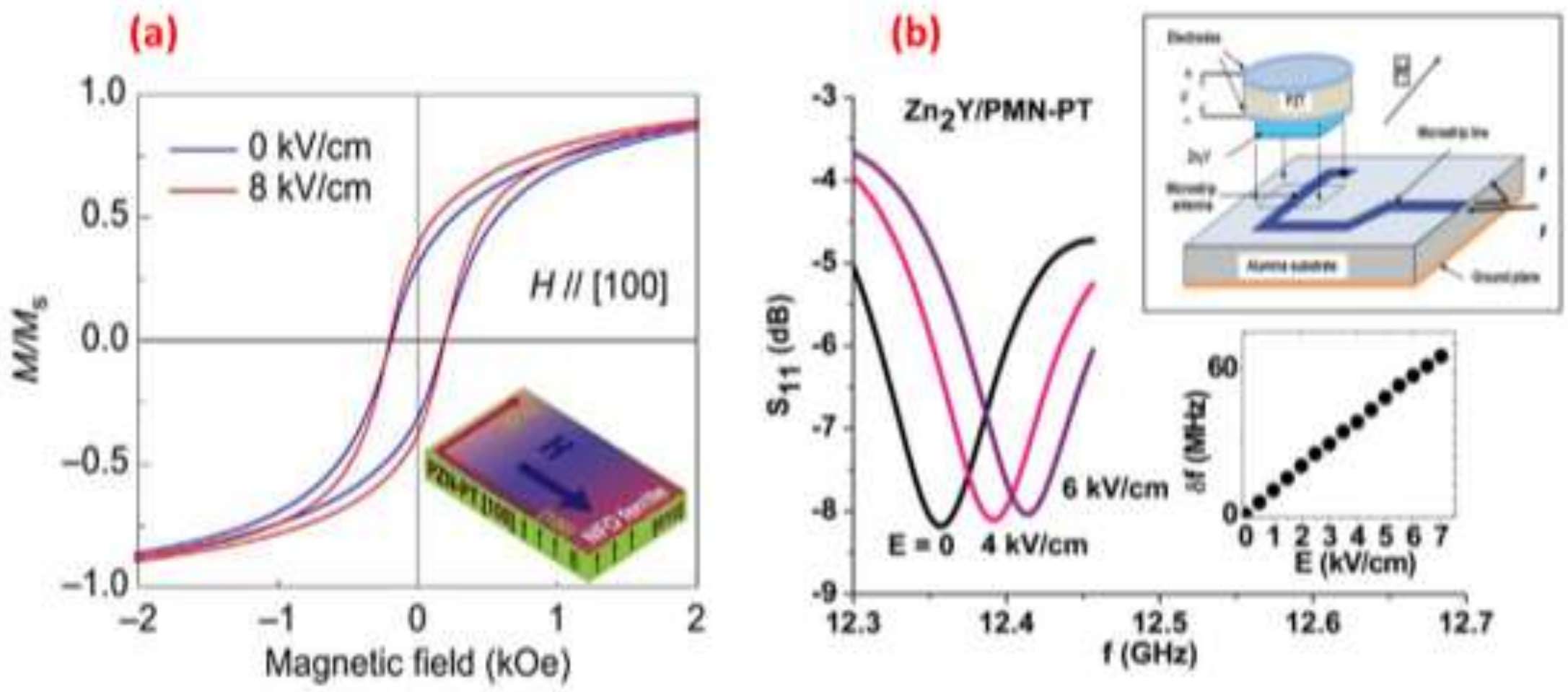

***Fig. 8.*** *(a) Magnetization M vs H for E =0 and 8 kV/cm for a thin film of nickel ferrite on $Pb(Mg_{1/3}Nb_{2/3})O_3$-$PbTiO_3$ (PMN-PT)* [100]*, (b) data showing tuning of FMR in bilayer of single crystal Y type hexagonal ferrite and PMN-PT; insets show strip-line device used for measurements and shift in FMR frequency versus E* [100]*.*

The FMR method relies on the magnetic resonance frequency's strain dependence, whose magnitude is determined by the piezoelectric and magnetostrictive constants. The mechanically induced strain, resulting from an applied electric field, manifests as an internal magnetic field within the composite, causing a shift $\delta H$ in the FMR field (Fig. 8b) [100]. This shift $\delta H$ is significantly influenced by various factors including sample magnetization, magnetostriction, and piezoelectric coefficient d [58].

Additional methods for measuring ME coupling in nanostructured samples include techniques such as piezo-response force microscopy (PFM), magnetic force microscopy (MFM), and ferromagnetic resonance (FMR) performed under an applied electric field using a scanning microwave microscope (SMM) [99,101–104].

## III Characterization of Candidate ME Materials

The ME effect is a complex phenomenon that involves the interaction of magnetization and polarization. To acquire a thorough understanding of the ME effect, it is necessary to explore the underlying mechanisms that contribute to coupling. The piezoelectric effect, for example, produces an electric potential in a material as a result of mechanical strain. Another mechanism that adds to the ME effect is the magnetostrictive or piezomagnetic effect, which is the production of a magnetic field in a material due to mechanical stress. In this section of the review, a detailed overview of magnetism and ferroelectricity, including fundamental principles, origins, and potential applications will be provided. It will also give a brief recent overview of the literature on piezoelectric and magnetostrictive materials.

## 1. Piezoelectric effect

Materials classified as piezoelectric materials can generate an electric potential when exposed to an applied mechanical stress (direct piezoelectric effect), or generate a mechanical movement when subjected to an electric field (converse piezoelectric effect). The direct piezoelectric effect is the most common effect observed in piezoelectric materials, where the application of a mechanical stress causes a crystal structure deformation that separates charges and generates an electric potential [105]. The piezoelectric effect was firstly discovered by the Jacques and Pierre Curie brothers, in 1880 [106]. They observed that an electrical charge accumulated in the material in inorganic crystals such as tourmaline, quartz, topaz and Rochelle salt when mechanical stress was applied and the generated voltage was proportional to the mechanical stress (Fig. 9a). The converse piezoelectric effect was also experimentally observed, by Gabriel Lippmann in 1881, in single crystals with acentric symmetry, where an external electric field generated a mechanical response in the crystal (Fig. 9b).

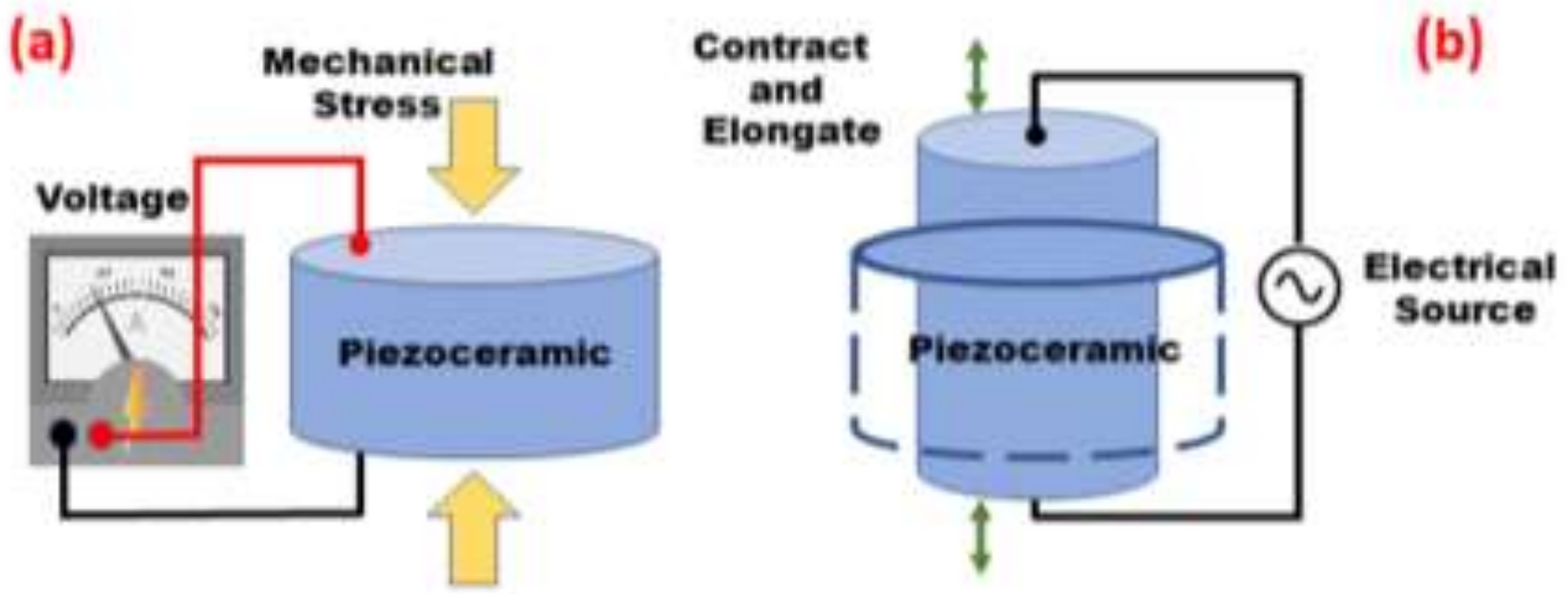

***Fig. 9.*** *Schematic diagram illustrating the piezoelectric effect. (a) Direct piezoelectric effect, (b) converse piezoelectric effect (After [107]).*

### a. Origin and related parameters

The chemical origin of piezoelectricity lies in the induced changes in the crystal structures of piezoelectric materials at the atomic level. When these materials are subjected to stress or pressure, their crystal structure deforms, leading to a change in the electrical current across the material. This deformation of the regular atomic pattern within the solid-state lattice produces the piezoelectric effect (Fig. 10)[105].

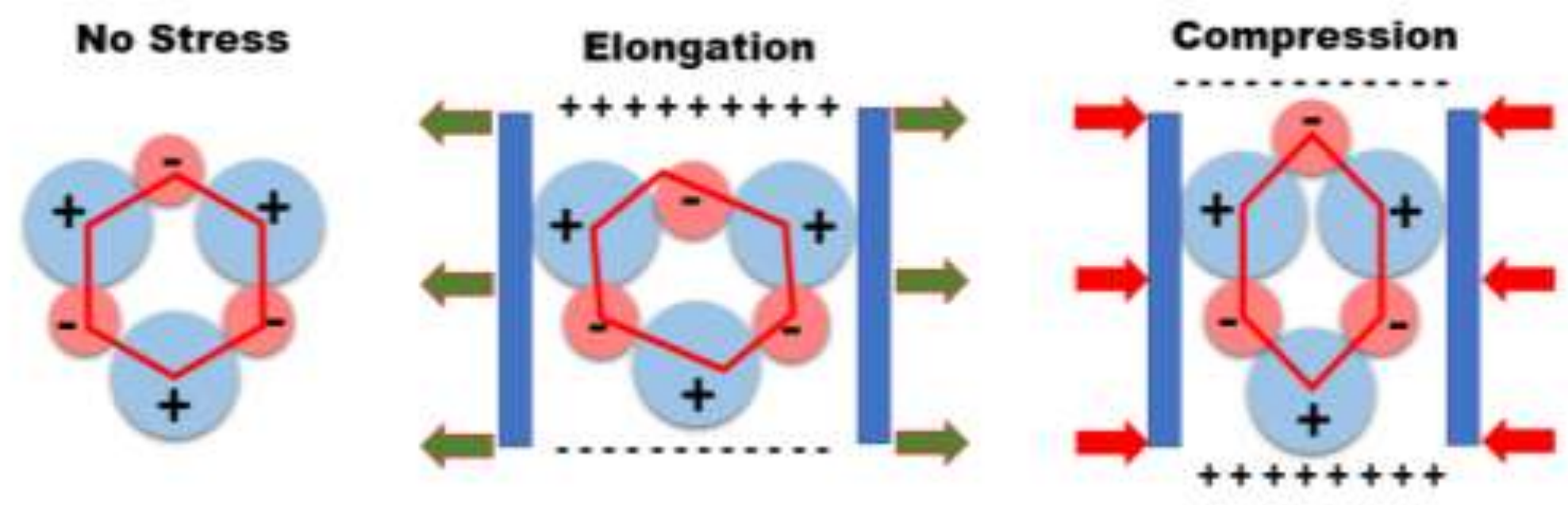

***Fig. 10.*** *Piezoelectricity at the atomic scale (After [107]).*

The piezoelectric effect occurs in most non-centrosymmetric crystals, where electric dipole moments are induced by stress. Of the 32 crystal classes, 21 are non-centrosymmetric, and all but one exhibit piezoelectricity. Pyroelectric materials must be acentric with a unique polar axis, while ferroelectric materials display reversible polarization under an external electric field. Inversion symmetry limits piezoelectricity [105]. Figures 11a and 11b summarize these properties and their interrelationships.

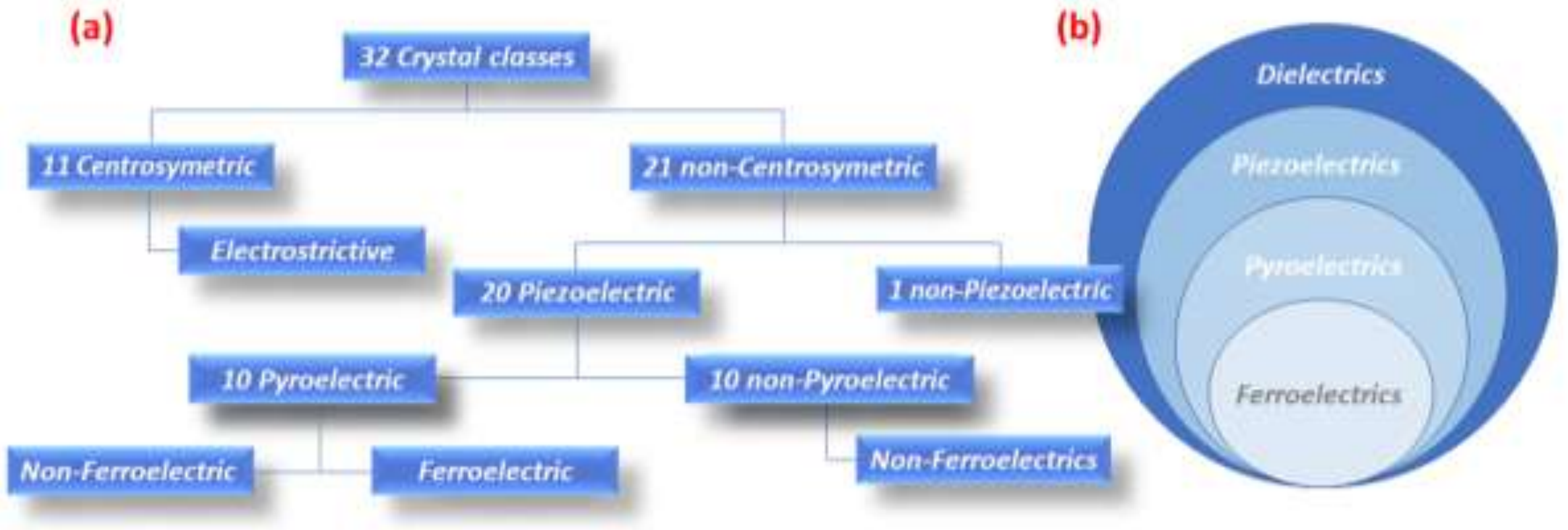

***Fig. 11.*** *(a) Schematic representation of piezoelectricity, pyroelectricity, and ferroelectricity based on crystal symmetry (after [108]), (b) A Venn diagram depicting how ferroelectrics range into the various dielectric material classes (after [109]).*

The piezoelectric effect demonstrates the relationships between mechanical variables (stress, X, and strain, S) and electrical variables (electric field, E, and electric displacement, D). Based on the selection of independent variables, the most commonly used piezoelectric coefficient written in the matrix form as $d_{ij}$. It represents the ratio of the surface charge per unit of mechanical stress applied for the direct effect, mostly used in sensors, according to the equation 8. For the actuators, it is given as the ratio of the deformation to the electric field for the converse effect (Eq. 9) [110].

$$d_{ij} = \frac{D_i}{\sigma_j} \quad \text{(Eq. 8)}$$

$$d^*_{ij} = \frac{S_j}{E_i} \quad \text{(Eq. 9)}$$

Here, $D_i$, $\sigma_j$, $S_j$, $E_i$ are dielectric displacement, applied stress, strain developed, and applied electric field. The coefficients are represented utilizing Voigt notation to denote the polarization direction and the mechanical stress direction along the *i* and *j* axes, respectively.

**b. Applications of piezoelectric materials**

Piezoelectric materials have found extensive applications in various fields, including actuators, resonators, sensors, transformers, capacitors, and transducers [111–113]. The first use of piezoelectric materials dates back to World War I, when they were used in an ultrasonic submarine detector employing a mosaic of tiny quartz crystals connected between two steel plates as a transducer [111]. Later discoveries resulted in the use of piezoelectric devices in resonating or non-resonating modes, such as signal filters, microphones, and ultrasonic transducers. However, the low performance of the materials at that time limited the economic viability of most devices. Thanks to advancements in the science and technology of piezoelectric materials, new high-performance materials have been discovered and developed. This progress has resulted in a significant commercial market for piezoelectric products, ranging from everyday use to specialized devices such as energy harvesters, sensors, actuators, ultrasonic sensors, airbag sensors, ceramic filters, resonators, buzzers, and transformers [114–117]. Figure 12 summaries the most important fields of piezoelectric materials applications.

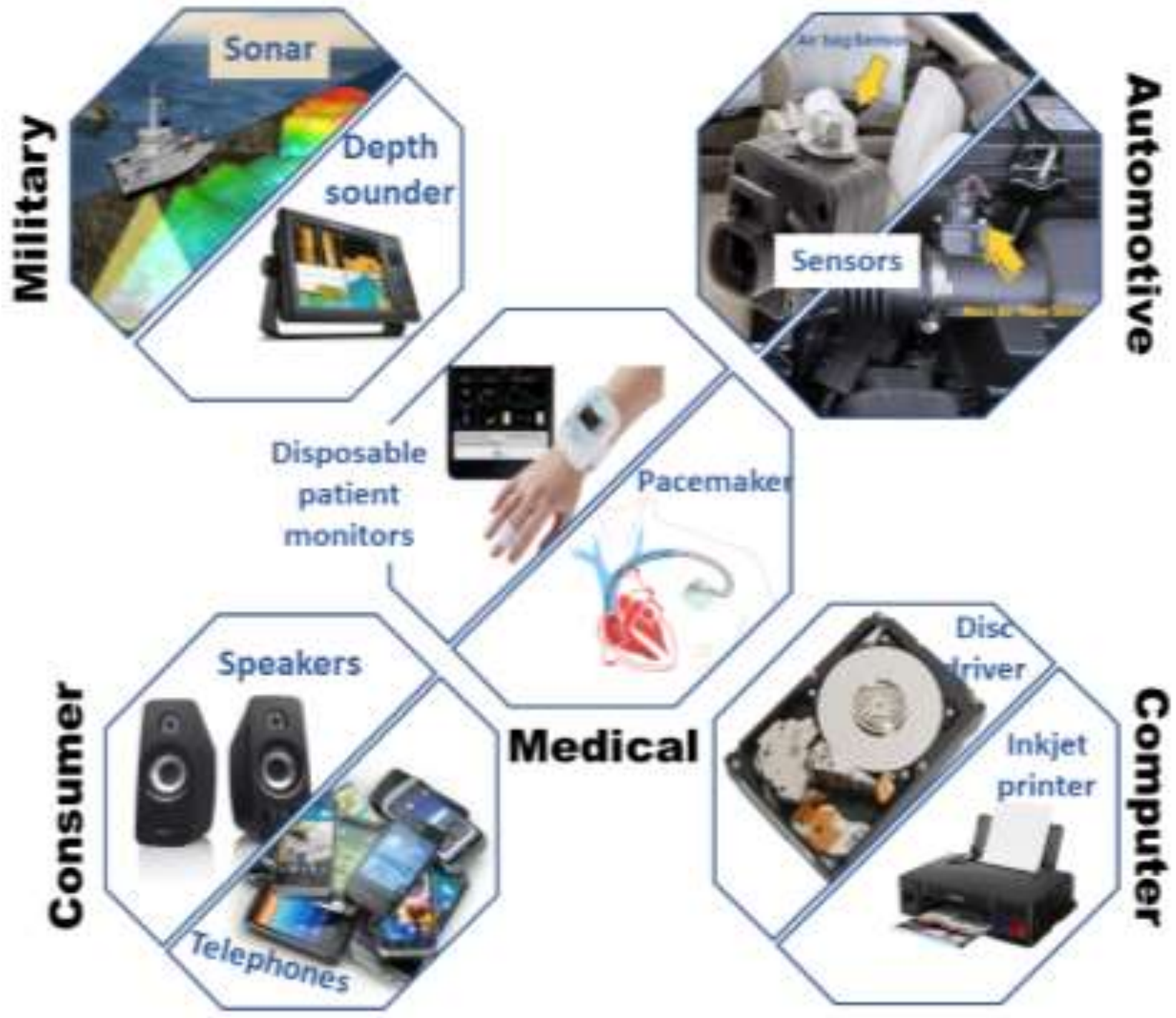

***Fig. 12.*** *Applications of piezoelectric materials.*

### c. Progress on Piezoelectric Materials

Following the discovery of barium titanate ($BaTiO_3$) ceramics in 1946, the first discovery in piezoelectricity was made, leading to a fast advancement in piezoelectric materials [118]. Jaffe et al. developed a series of lead zirconate titanium (PZT) ceramics in 1954, which had higher piezoelectricity than $BaTiO_3$ [119].

To date, a variety of piezoelectric ceramics have been used, including PZT, and PMN–PT (lead magnesium niobite – lead titanate) [120–123]. However, it is clear from the literature that PZT is the most commonly used piezoelectric material since it sinters easily, has a high electromechanical coupling, and doesn't react with other components. Motivated by environmental concerns, because of the toxicity of lead oxide, recent researches have been focused on "lead-free materials" [124–127].

Three types of lead-free piezoelectric ceramics exist: tungsten bronze, bismuth layered structures, and perovskite. The $ABO_3$-type family of perovskites is one of the most well-known classes of the lead-free piezoelectric materials, including $K_{0.5}Na_{0.5}NbO_3$ (KNN)[128], $Bi_{0.5}Na_{0.5}TiO_3$ (BNT) [129], $BaTiO_3$ (BT) [125,130], and $BiFeO_3$ (BFO) [131–133].

A high piezoelectricity of $d_{33} \approx 416$ pC $N^{-1}$ is measured in textured KNN-based ceramics due to the occurrence of an orthorhombic-tetragonal (O-T) phase boundary [134]. Thereafter, an ultrahigh piezoelectric property with $d_{33} \approx 700$ pC $N^{-1}$ and $d^*_{33} \approx 980$ $pmV^{-1}$ were achieved in highly textured (K, Na) $NbO_3$-based ceramics as reported by Peng Li [135]. In 2017, a high piezoelectric constant ($d_{33}$ = 525 pC $N^{-1}$) was obtained in (1-x)($K_{1-y}Na_y$)($Nb_{1-z}Sb_z$)$O_3$–x$Bi_{0.5}$($Na_{1-w}K_w$)$_{0.5}$$HfO_3$ (KNNS–BNKH), where a rhombohedral (R) - tetragonal (T) (R–T) phase boundary was developed by optimizing the KNNS–BNKH composition [136]. Despite the strong piezoelectricity properties reported in KNN-based ceramics, they are still inferior to those of PZT ceramics, which is inconvenient for practical applications. As a result, there is still a lot of effort to be done to improve the piezoelectricity of KNN-based materials.

Bismuth sodium titanate (BNT) has a low piezoelectricity (73-95 pC $N^{-1}$). As a result, to increase the piezoelectricity response of these materials, chemical changes such as ion substitution or solid solution are used to create phase boundaries [137,138].

Note that BNT ceramics have two types of phase boundaries. The first is defined as the transition between ferroelectric R and T phases, while the second is defined as the transition between relaxor (nonpolar) and ferroelectric (polar) phases [139]. In this respect, many studies have been conducted to increase the piezoelectricity of BNT materials by chemical composition modifications. The addition of a second material to form of binary systems is a viable approach. Takenaka et al. proposed a BNT-BT binary system for the first time in 1991, and discovered an improvement of $d_{33}$ of around 125 pC $N^{-1}$ in (1-x)BNT-xBT (x=0.06) due to the formation of R-T phase boundary [140]. This finding opened the way for further research into BNT-BT, providing some interesting results in terms of properties and physical processes. For instance, Zhang et al. studied the temperature-dependent electrical properties of BNT-BT ceramics and found that developing antiferroelectric order causes the largest unipolar strain (~0.42%) [141]. In addition, according to Simos et al., the structural transition induced by an electric field is the result of the significant increase in strain in BNT-BT ceramics. They also suggested that the reversible nature of the phase change caused the significant recoverable strain at high temperatures. Also, they attributed the significant recoverable strain at high temperatures to the reversible nature of the phase transition [142,143]. Similarly to PZT ceramics, 0.93BNT-0.07BT ceramics were found to have an intermediate monoclinic phase at the morphotropic phase boundary (MPB), allowing polarization rotation and improved piezoelectric properties [144]. To provide further evidence proving the efficacy of phase boundaries in enhancing the piezoelectric properties of BNT, various investigations explored the substitution of Ti/Ba ions with Zr [145], Hf [146], Ca [147], and Sr [148]. This substitution has generated a significant piezoelectric response in the phase boundary region. Many binary systems have also been explored, including BNT combined with $Bi_{0.5}K_{0.5}TiO_3$ [149],

$K_{0.5}Na_{0.5}NbO_3$ [150], $BaSnO_3$ [151], $Bi(Zn_{0.5}Ti_{0.5})O_3$ [152], $Bi(Mg_{0.5}Ti_{0.5})O_3$ [153], and $Bi(Al_{0.5}Ga_{0.5})O_3$ [154]. To further enhance the piezoelectricity of BNT, ternary systems have also been investigated, such as BNT-BT-KNN and BNT-BKT-KNN [155,156]. Notably, the ternary system $0.854Bi_{1/2}Na_{1/2}TiO_3$-$0.12Bi_{1/2}K_{1/2}TiO_3$-0.026 $BaTiO_3$ with an R-T phase boundary exhibited a high piezoelectricity value of about 295 pC $N^{-1}$ [157]. Nevertheless, despite the efforts to improve the BNT's functionalities, its piezoelectric response remains insufficient for integration into real-world applications.

Owing to its high $T_c$ and good electrical properties, BFO-based ceramics are considered promising materials in high-temperature applications [158–161]. However, this pure BFO material has no effective phase boundary that would improve its piezoelectric properties[162]. Hence, Bi and/or Fe sites substitutions are a beneficial way to improve the piezoelectric properties by constructing a temperature-independent phase boundary. The Bi site can be replaced by a variety of chemical elements, such as: Ca, Sm, La, Nd, Dy, Eu, Y, Ce, and Ho [131,163–168]. For instance, Zheng et al. observed an improvement in the piezoelectric coefficient ($d_{33} \approx 50$ pC $N^{-1}$) through the substitution with Sm and La in BFO-based ceramics [169]. This enhancement was mainly attributed to the suppression of impure phases and the reduction in leakage current. For Fe site, Sc-doped BFO ceramics show enhanced electrical properties of $d_{33} \approx 46$ pC $N^{-1}$ as reported by Lv et al. [170]. Furthermore, it was discovered that the BFO's piezoelectric response may be further enhanced by an adequate substitution in both Bi and Fe sites. For example, Troyanchuk et al. reported an increase of $d_{33} \approx 120$ pC $N^{-1}$ in $Bi_{0.82}Ca_{0.18}Fe_{0.91}Nb_{0.09}O_3$ [171].

In binary systems, the presence of phase boundaries and higher resistivity can significantly boost the piezoelectric properties compared to BFO ceramics with ion substitution. For example, in (1-x)$BiFeO_3$–x$BaTiO_3$ (BFO–BT) ceramics composites, an enhanced of piezoelectric coefficient $d_{33}$ from 76 to 274 pC $N^{-1}$ was observed due to the involvement of phase boundaries (R-T) [172]. By adding a third component, the ternary system enhanced the electrical properties of BFO–BT ceramics. For example, in ceramics with an R-T phase boundary, BFO–BT–BZT (BZT: $BaZr_{1-x}Ti_xO_3$) showed a large $d_{33}$ of 324 pC $N^{-1}$[173].

As the earliest-discovered polycrystalline ceramics, the BT-based ceramics exhibited much higher piezoelectricity than other lead-free piezoceramics [174]. As was previously mentioned, materials engineering is a useful technique for enhancing the electrical properties of piezoelectric ceramics. It has been confirmed previously that phase transitions in BT ceramics are influenced by the addition of some elements such as Zr, Hf, Sn, or Ca. Table 1 summarizes the effects of these doping elements on the phase transition temperatures of BT ceramics[175–179]. From these results, its revealed that an appropriate substitution can adjust $T_{R\text{-}O}$ and/or $T_{O\text{-}T}$ to near RT, where distinct phase boundaries of BT may be constructed. It is worthy to mention that the R-O phase boundary was rarely taken into consideration because of the low piezoelectricity, which generally have far inferior piezoelectricity than those with R-T and R-O-T phase boundaries. Consequently, it is possible to significantly increase the piezoelectricity of BT-based ceramics by increasing the coexistence of R and T phases.

***Table 1:*** *Effect of some typical substitutions on phase transition of BT-based materials (↓: decrease, ↑: increase, ↔: basically unchanged).*

| **$BaTiO_3$** | **Chemical elements** | **Sites** | **Ionic radii (nm)** | **$T_{R\text{-}O}$** | **$T_{O\text{-}T}$** | **$T_C$** |
|---|---|---|---|---|---|---|
| **A-site $Ba^{2+}$: 0.135** | $Ca^{2+}$ | A | 0.100 | ↓ | ↓ | ↔ |
| | $Sr^{2+}$ | A | 0.118 | ↓ | ↓ | ↓ |
| **B-site $Ti^{4+}$: 0.06** | $Zr^{4+}$ | B | 0.072 | ↑ | ↑ | ↓ |
| | $Sn^{4+}$ | B | 0.069 | ↑ | ↑ | ↓ |
| | $Hf^{4+}$ | B | 0.071 | ↑ | ↑ | ↓ |

The (Ba, Ca)(Ti, M)$O_3$ (M = Zr, Sn) ceramics with R-T or R-O-T phase boundaries have been studied for their enhanced electrical properties. These materials have shown potential for improved electrical performance and temperature stability near RT [84,179,180]. In this regard, lead-free $Ba(Ti_{0.8}Zr_{0.2})O_3$–$(Ba_{0.7}Ca_{0.3})TiO_3$ (BCZT) ceramics with an R-T phase boundary exhibit a large piezoelectric effect ($d_{33} \approx 620$ pC $N^{-1}$) as reported by Liu and Ren [181]. The high piezoelectric constant was primarily attributed to the nearly vanishing polarization anisotropy and enhanced polarization rotation between the R and T states due to the involvement of MPB composition near the tricritical triple point. Many researchers have been involved in synthesizing either the same MPB or modified-BCZT ceramics to achieve the improved/similar values of $d_{33}$ reported by Liu and Ren or greater/close to commercial PZT-5H ceramics. By taking advantage of R-T phase boundary, Ehmke et al. obtained a giant strain value of $d^*_{33}$ ~ 1310 pm $V^{-1}$ in $0.55Ba(Zr_{0.2}Ti_{0.8})O_3$–$0.45(Ba_{0.7}Ca_{0.3})TiO_3$ (0.55BZT–0.45BCT) ceramic [182]. On the other hand, the presence of the orthorhombic intermediate phase can also yield significant electrical properties. For 0.5BZT–0.5BCT, Zhang et al. found a maximum of $d_{33}$ near the O-T phase boundary due to the facile polarization rotation, larger lattice softening, and reduced anisotropic energy [183]. As a result, modifying phase compositions is critical for modifying the electrical properties of such ceramics. In another context, the properties of lead-free ceramics can be modified through various process conditions. For example, by optimizing poling conditions, BCZT ceramics have achieved a substantial piezoelectric coefficient $d_{33}$ of 637 pC $N^{-1}$, a significant electromechanical coupling coefficient ($k_p$) of 0.596, and a high piezoelectric voltage coefficient ($g_{33}$) of 29 mV m $N^{-1}$ [184]. In 2015, Klara et al. investigated the impact of synthesis methods on piezoelectric properties and observed that $Ba_{0.85}Ca_{0.15}Zr_{0.1}Ti_{0.9}O_3$ ceramics elaborated by sol-gel method and sintered at 1425°C exhibited the highest $d_{33}$ value of 410.8 ± 13.2 pC $N^{-1}$ [185]. Doping effect was reported by Kim et al. and observed an excellent piezoelectric property ($d_{33}$ ~ 623 pC $N^{-1}$, and $k_p$ ~ 51%) for the $Sb_2O_3$-doped BCZT ceramics [186]. Recently, Shi et al. discussed the impact of sintering temperature and achieved an enhanced piezoelectric coefficient, $d_{33}$ ∼ 529 pC $N^{-1}$, for the BCZT sample using two-step sintering [187]. To the best of our knowledge, the $[0\ 0\ 1]_c$ grain-oriented $(Ba_{0.94}Ca_{0.06})(Ti_{0.95}Zr_{0.05})O_3$ ceramics with R-O-T phase boundaries exhibit the highest piezoelectricity observed in BCZT ceramics to date ($d_{33}$ = 755 pC $N^{-1}$) [188]. Accordingly, creating an R-T or R-O-T phase boundary and adjusting the conditions of processing are extremely effective ways to enhance the electrical characteristics of $(Ba,Ca)(Ti,Zr)O_3$ ceramics.

$(Ba,Ca)(Ti,Sn)O_3$, another significant system like BCZT, exhibits good electrical properties with R-T or R-O-T phase boundary. For instance, $Li_2O$-modified $(Ba_{0.95}Ca_{0.05})(Ti_{0.90}Sn_{0.10})O_3$ ceramics with an R-T phase boundary, demonstrated a substantial $d_{33}$ of 578 pC $N^{-1}$ [189]. Moreover, the presence of an R-O-T phase boundary in $0.55(Ba_{0.9}Ca_{0.1})TiO_3$–$0.45Ba(Sn_{0.2}Ti_{0.8})O_3$ ceramics results in an even higher $d_{33}$ of 630 pC $N^{-1}$ and a $k_p$ of 52% [190]. Additionally, $(Ba_{0.95}Ca_{0.05})(Ti_{0.91}Sn_{0.09})O_3$ with a pseudo cubic Pc-O phase boundary exhibited a remarkable $d_{33}$ of 670 pC $N^{-1}$ [191].

Generally, ceramics with R-T or R-O-T phase boundaries exhibit better piezoelectric properties than those with O-T or R-O phase boundaries. Since the R-T or R-O-T phase boundary contributes differently to the polarization rotation anisotropy and the domain-wall, it can significantly increase the piezoelectric response.

## 1. Piezomagnetic effect and magnetostrictive effect

The magnetostrictive effect, first described by James Prescott Joule in 1842, refers to the property of ferromagnetic materials (iron) to change their shape or dimensions in response to a magnetic field. This effect allows magnetostrictive materials to convert electromagnetic energy into mechanical energy and vice versa (Fig. 13a). The material strains and elongates when a magnetic field is applied as a result, the material's molecular dipoles and magnetic field boundaries rotate in order to align with the field (Joule effect). Conversely, applying stress to a magnetostrictive material changes its magnetization (Villari Effect) [192].

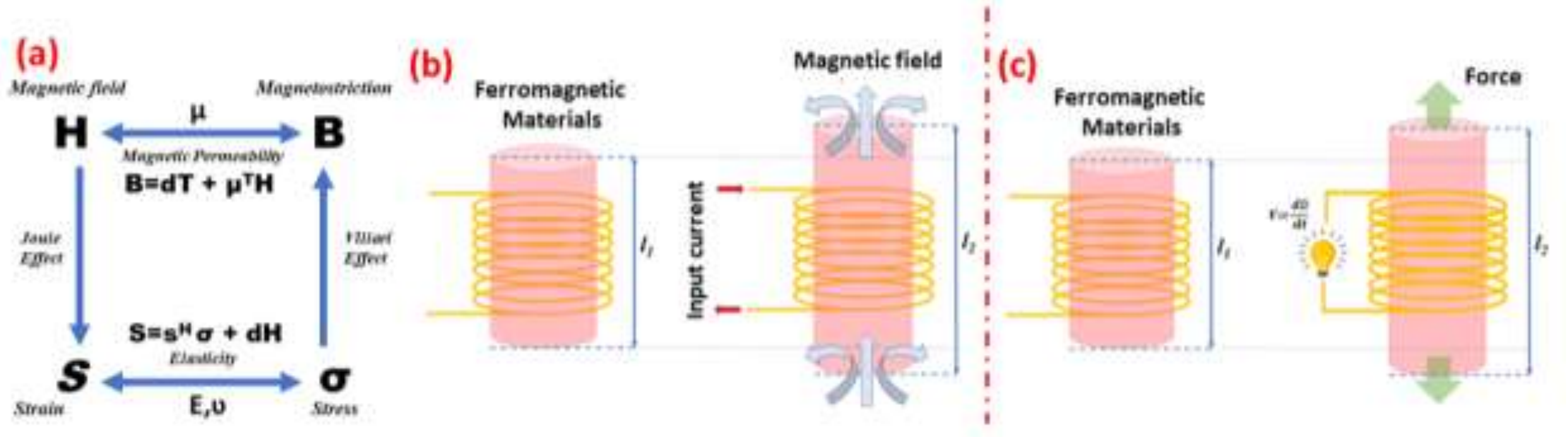

***Fig. 13.*** *(a) Magnetostrictive effect, (b) Joule Effect (Direct Magnetostrictive Effect), (c) Villari Effect (Inverse Magnetostrictive Effect) (After [192]).*

The magnetostrictive effect is quantified by magnetostrictive strain λ, defined as the ratio of the change in length Δl to the original length l of the material ($\lambda = \frac{\Delta l}{l}$). λ is generally expressed as $\times 10^{-6}$ or ppm and can be positive or negative depending on whether the magnetic material elongates or contracts [193]. This bidirectional energy exchange property makes magnetostrictive materials useful for various applications, including sensors, transformers, and actuators [194]. Figure 13 illustrates both (b) direct and (c) inverse magnetostrictive effects.

### a. Origin of magnetostriction

According to the minimal free energy principle, alterations in material's magnetized state cause changes in its length or volume, which are necessary to reach the minimum total energy in the system [195]. Magnetostrictive phenomena in magnetic materials arise from various mechanisms. Spontaneous magnetostriction, driven by exchange forces, occurs in single-domain crystals below the Curie temperature. Here, spontaneous magnetization aligns atomic magnetic moments, causing nuclear separation and resulting in magnetostrictive strain [196]. Field-induced magnetostriction involves the movement of magnetic domain walls and the rotation of internal domains in ferromagnetic materials. This process redistributes energy to achieve the system's lowest energy state, altering magnetoelastic properties and causing noticeable changes in the material's size [197]. Furthermore, the shape effect is another way magnetostriction works, connected to the magnetic material's shape and demagnetization energy (1/2N$M_S^2$V). To minimize demagnetization energy, it is necessary to decrease the sample volume (V) and lower the demagnetization factor (N) by elongating it along the magnetization direction. The shape effect due to demagnetization energy is comparatively smaller than other magnetostrictive effects [198,199].

Giant magnetostriction, predominantly observed in rare earth metals, alloys, and intermetallic compounds, arises from the unfilled 4f electron orbitals in rare earth ions. The strong anisotropy of these orbitals induces significant lattice distortions along specific directions when spontaneously magnetized, leading to pronounced magnetostrictive effects [200–202].

### b. Applications of magnetostrictive materials

Magnetostrictive materials are used to convert electromagnetic energy to mechanical energy and vice versa. This phenomenon may be exploited to make sensors that detect magnetic fields or sense forces. The imparted magnetic field or force would generate a measurable strain in the material [201].

Transformers employ magnetostrictive materials and Faraday's law to turn magnetic fields into electromotive forces. With this process, the shift in magnetic flux is converted into an electromotive force (EMF) in the transformer. This effect is utilized to boost or reduce alternating current voltages as well as to transfer them from one circuit to another. In response to a changing magnetic field (wave), a magnetostrictive material spins its molecular dipoles in phase with the frequency. The largest change in length occurs twice per magnetic field period. This produces a buzzing sound, which is typical in transformers [203,204].

To generate vibrations, a changing magnetic field can be utilized in combination with magnetostrictive materials. Medical devices and industrial vibrators, ultrasonic cleaning equipment, underwater sonar, vibration or noise control systems, and a variety of other applications make use of such materials. A mechanical lever can be used to enhance the amplitude of vibration. On the other hand, magnetostrictive transducers may be used to transmit ultrasonic energy into other materials [192,205,206].

### c. Progress on Magnetostrictive Materials

Since their discovery in the 1960s, rare earth (RE) based alloys magnetostrictive materials, such as Terfenol-D, Galfenol, and Samfenol-D, have been considered promising options due to their large magnetostriction. These materials exhibit magnetostrictive effect that are hundreds of times greater than those of Fe, Ni, and Co [207]. However, recent research claims that magnetostrictive oxides can serve as an alternative to expensive alloy based rare earth magnetostrictive materials [208]. The development of ceramic oxide-based magnetostrictive materials is currently dealing with a greater attention to address the high cost, eddy current effects, chemical and high temperature stability, and other issues related to alloy-based magnetostrictive materials [209]. At low temperatures, several of the metal oxides perovskites, especially manganates, exhibit massive magnetostriction; however, at ambient temperature, this magnetostriction disappears [210].

Cobalt ferrite, $CoFe_2O_4$ (CFO), belongs to the family of spinel-type ferrites, exhibits remarkable negative magnetostriction at room temperature. Its magnitude in single crystal form is estimated to be around -600 ppm in the [100] crystallographic direction [211]. Depending on the synthesis method, its magnitude (in polycrystalline sintered materials) is known to vary from -100 ppm to - 400 ppm. Many efforts have been made in recent years to improve the strain and strain sensitivity of sintered cobalt ferrites by modifying their chemical composition, such as the substitution of magnetic (Ni, Mn, Cr) or non-magnetic elements (Mg, Cu, Zn, Al, Zr, In, Ga, Ti, Ge) for Fe or Co in CFO structure [194,201,202,209,212–219]. Although, magnetostriction of the Cu-Mn-co-doped CFO system is comparable to the doped-Mn CFO system, an appreciable increase in piezomagnetic coefficient $(d\lambda/dH)_{max}$ as high as 0.2017 ppm $Oe^{-1}$ was found in the co-doped CFO by Siva et al [215]. Table 2 gathers the magnetostrictive effect in ferrite-based materials.

***Table 2:*** *Magnetostrictive effect in ferrite-based materials*

| Sample composition | $-\lambda_s$ (ppm) | $-(d\lambda/dH)_{max}$ (ppm $Oe^{-1}$) | References |
|---|---|---|---|
| $CoFe_2O_4$ | 400 | 0.002 | 194 |
| $MgFe_2O_4$ | 6 | - | 12 |
| $MnFe_2O_4$ | 5 | - | 12 |
| $NiFe_2O_4$ | 35 | - | 220 |
| $NiGa_{0.5}Fe_{1.5}O_4$ | 22 | - | 221 |
| $CuFe_2O_4$ | 9 | - | 12 |
| $CoFe_{1.8}Mn_{0.2}O_4$ | 117 | - | 218 |
| $Co_{1.1}Fe_{1.85}Nb_{0.05}O_4$ | 123 | - | 214 |
| $CoFe_{1.9}Zn_{0.1}O_4$ | 148 | 0.105 | 213 |
| $Co_{0.7}Mn_{0.3}Fe_{1.9}Dy_{0.1}O_4$ | 94 | 0.075 | 214 |
| $Co_{0.95}Cu_{0.05}Fe_2O_4$ | 170 | 0.0016 | [217] |
| $CoBi_{0.2}Fe_{1.8}O_4$ | 196 | 0.0015 | [216] |
| $CoAl_{0.1}Fe_{1.9}O_4$ | 230 | 0.247 | [219] |
| $Fe_2O_3$ | - 40 | - | [12] |
| $SmFe_5O_{12}$ | - 3.3 | - | [12] |
| $EuFe_5O_{12}$ | - 9.48 | - | [12] |

## IV Magnetoelectric Materials: Types and Connectivities

Single-phase multiferroic materials have been limited in number, primarily due to the inherent contradiction between the conventional mechanism in ferroelectric oxides, which necessitates empty d-orbitals, and the development of magnetic moments, arising from partially filled d-orbitals. An alternative approach involves the fabrication of artificial multiferroics, wherein two distinct compounds – one ferromagnetic and the other ferroelectric – are utilized. The objective of this methodology is to shape materials that manifest the properties of the parent compounds while establishing a coupling between them. However, it is important to note that the cross coupling is typically achieved indirectly through strain (comprising magnetostriction in addition to electrostriction and/or piezoelectricity), rather than through a direct interaction between polarization and magnetization (*P*-*M*) in the material [8,10,222].

### 1. Single phase

In transition metal or rare-earth ions partly filled ***d*** or ***f*** shells, there are localized electrons that have a corresponding localized spin, or magnetic moment. This is the microscopic origin of magnetism, which is essentially the same in all magnets. Magnetic order is produced by exchange interactions between the localized moments [222]. In the case of ferroelectrics, the scenario is very different. There are several various ferroelectricity sources, and as a result, there are different types of multiferroics.

Single-phase multiferroic materials are characterized as chemically isotropic, homogeneous compositions whereby electric and magnetic order states coexist at any given place or region in the material. They can have various kinds of magnetic ordering like collinear, spiral or frustrated spin structure [6]. In this type of multiferroics, symmetry breakdown at magnetic surfaces induces a weak electric polarization during the magnetic reordering or magnetic phase transition. However, due to low antiferromagnetic transition temperature in single phase multiferroics, this effect is either very weak or only detected at low temperatures [223,224].

In general, there are two categories of multiferroics single phase: ***Type-I single-phase*** multiferroics, in which magnetic and electric phases coexist within the same compound, but they have different microscopic origins and are broadly independent of each other, however there is some coupling between them. Typically, ferroelectricity appears at temperatures greater than magnetism in these materials, and the spontaneous polarization $P^s$ is frequently rather significant (on the order of 10 - 100 μC $cm^{-2}$) [225]. ***Type-II single-phase*** multiferroics are materials in which ferroelectric order is induced by magnetism, implying a very strong magneto-electric coupling. However, the polarization in this material is significantly lower (~ $10^{-2}$ μC $cm^{-2}$) [226].

### a. Type-I multiferroics

#### i. Ferroelectricity due to lone pairs

The origin of ferroelectricity in BFO, for example, arises from $Bi^{3+}$ ions. According to their electronic configuration, these ions, called "lone pairs", have two outer 6s electrons that do not participate in chemical bonds. However, they possess high polarizability, which is essential for ferroelectricity (Fig. 14a) [60,227].

#### ii. Ferroelectricity due to charge ordering

This mechanism is commonly observed in compounds based on transition metals, especially those featuring metals with mixed valences and experiencing geometrical or magnetic frustration. As shown in Figure 14b, improper ferroelectricity occurs when active sites and bonds are not equivalent after charge ordering, which is characterized by a lack of ionic displacement. $LuFe_2O_4$ is a prominent example of this phenomena, where charge ordering causes ferroelectricity to arise at 332 K. Due of the mixed valence of Fe ions on the triangular lattice, the structure's electronic arrangement is disrupted, which causes charge frustration and mild ferromagnetism [228].

#### iii. "Geometric" ferroelectricity

Material instability can be triggered by geometric constraints and the size effect, with steric effects playing a more significant role than the typical changes in chemical bonds. This steric-driven process results in an ionic shift, causing polar distortion and geometric ferroelectricity. A notable category of materials that defies the '$d^0$-ness' rule includes hexagonal manganites, also known as hexagonal perovskites ($RMnO_3$, where R=Y or small rare earths) [222]. Despite sharing an apparently similar $ABO_3$ formula, these systems exhibit distinct crystal and electronic structures. They exhibit ferroelectricity at high transition temperatures (900 – 1000

K) and ferromagnetism at very low Neel transition temperatures ($T_N \leq 120K$) [229]. In these compounds, ferroelectricity is almost an “accidental by-product” of the tendency to close packing. The tilting of the fixed $MnO_5$ block, with the Mn ion at the center, disrupts inversion symmetry and leads to ferroelectricity. The resulting dipole moments are predominantly formed by A–O pairs. As a consequence of the tilting of the $MnO_5$ block, A–O bonds form electric dipoles, generating two 'up' dipoles for every 'down' dipole. This configuration contributes to the material's ferroelectric nature, and when Mn spins order at lower temperatures, it also exhibits MF nature (Fig. 14c) [230].

**b. Type-II multiferroics**

This type of multiferroics exhibits ferroelectricity driven by magnetism, leading to strong coupling between magnetic and electric orders. Most rare-earth manganites falling into this class, such as $RMnO_3$ and $RMn_2O_5$ (where R represents rare earths), show a significant ME effect [67,231–233].

However, a notable drawback of these materials is the relatively low polarization they induce, typically in the order of nC $cm^{-2}$. This value is approximately 1000 times too small for effective discrimination by the sense amplifiers during a READ operation in a Ferroelectric Random-Access Memory (FeRAM), and this limitation persists even at temperatures significantly below room temperature. In terms of the mechanism underlying multiferroic behavior, type-2 multiferroics can be further classified into three types, (i) Symmetric spin exchange interaction, (ii) antisymmetric spin exchange interaction and (iii) spin ligand interaction (spin dependent p-d hybridization) (Fig. 14d) [222,234].

(i) Magnetism in materials is attributed to the alignment of electron spins, which give rise to intrinsic magnetic moments. The interaction between neighboring spins, denoted as $S_i$ and $S_j$, introduces strain and disrupts inversion symmetry, causing the electron spins to preferentially align perpendicular to each other. This leads to the formation of noncollinear and chiral magnetic structures, resulting in the appearance of polarization (P). This antisymmetric interaction is recognized as the Dzyaloshinskii–Moriya (DM) interaction.

(ii) This type of interaction is often referred to as the inverse DM interaction. In this mechanism, the polarization arises from the canted orientation of spin sites $S_i$ and $S_j$, where spins are tilted by a small angle about their respective axes instead of being exactly parallel. This polarization is attributed to the spin current between sites $i$ and $j$.

(iii) In contrast to the spin exchange interaction and inverse DM interaction, the p-d hybridization mechanism involves only one magnetic site coupled with a ligand ion to induce polarization. Within this mechanism, the d orbital of the magnetic ion and the p orbital of the ligand, separated by their spin states due to spin-orbit interaction, hybridize with each other. A notable characteristic of this mechanism is that the electric polarization is contingent upon the spin state of the hybridized orbital within the local cluster. Moreover, this hybridized orbital may influence magnetic anisotropy via spin-orbit interaction.

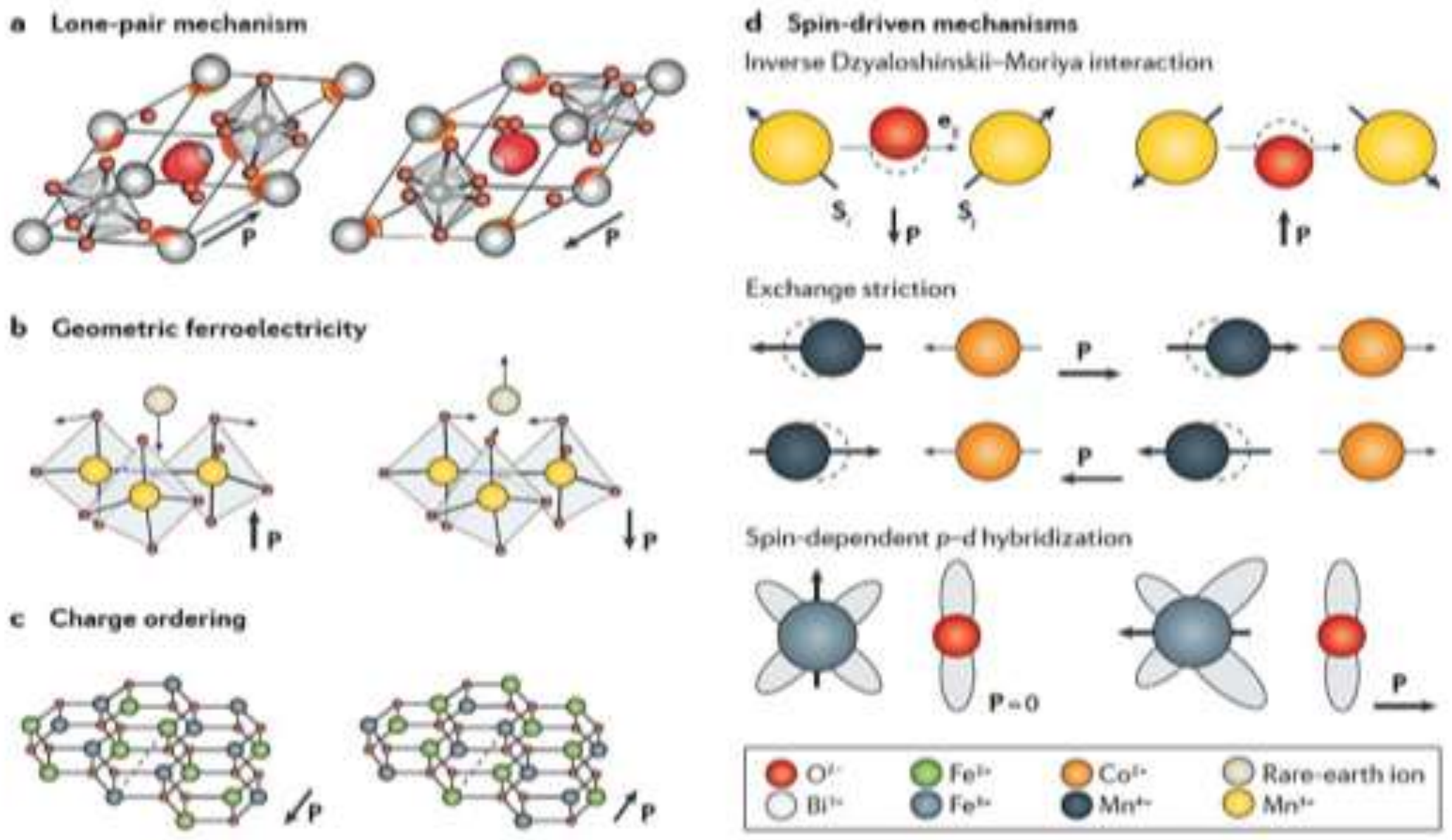

***Fig. 14.*** *(a) Lone-pair ferroelectricity in BFO. (b) Geometrically driven ferroelectricity in hexagonal (h-) $RMnO_3$ emerges from a tilt and deformation of $MnO_5$ bipyramids, which displace the rare-earth ions as indicated by the arrows, leading to a spontaneous polarization along the [001] axis. (c) Charge ordering in $LuFe_2O_3$ creates alternating layers with $Fe^{2+}/Fe^{3+}$ ratios of 2:1 and 1:2. This was argued to create a spontaneous polarization between the two layers, which is oriented parallel to the arrow. (d) Mechanisms for spin-induced ferroelectricity. Polar displacement is induced by antisymmetric spin exchange interactions (inverse Dzyaloshinskii–Moriya interaction; top panel). Ferroelectricity arises from symmetric spin exchange in $Ca_3CoMnO_6$ shown in the middle panel. Spin-driven modulations of the chemical bond between magnetic 3d orbitals and ligand 2p orbitals (indicated by grey clouds) yield a spontaneous polarization along the bond direction in delafossites, such as $CuFeO_2$ (Reprinted with permission from ref [222]).*

## 2. Artificial composite multiferroics

Even though extensive research has been conducted on single-phase multiferroic compounds, no material has yet demonstrated a significant and practically usable magnetoelectric coupling at room temperature. The absence of room temperature behavior and the weak magnetoelectric coupling make them unsuitable for various applications. Consequently, academics and researchers are driven to explore alternative multiferroic materials.

One innovative approach involves the development of "artificial" multiferroic composite materials. Multiferroic composites, by definition, are compounds wherein electric, magnetic, and elastic order states coexist. These orders are dimensionally separated from each other within the multiferroic composite material. As a result, the fabrication of multiferroic composites enables the creation of a diverse range of materials with the potential to optimize and control magnetoelectric coupling. As previously mentioned, the connectivity of ferroelectric and magnetic phases plays a crucial role in determining the overall magnetoelectric properties. The magnetoelectric coupling in such composites, as already discussed, is an indirect coupling, mediated via mechanical strain, between two active solids that individually exhibit magneto-elastic and electro-elastic couplings, respectively. This coupling, known as the product property of composite materials, can be tailored by the appropriate choice of phases with piezomagnetism (or magnetostriction) and piezoelectricity (or electrostriction) and their connectivity. Commonly cited connectivity types in the literature involve ferroelectric phases such as BT and its derivatives, $PbTiO_3$ (PT), $Pb(Zr,Ti)O_3$ (PZT), and BFO, and magnetic materials including CFO, $NiFe_2O_4$ (NFO), $La_{1-x}Sr_xMnO_3$ (LSMO), and $La_{1-x}Ca_xMnO_3$ (LCMO) [49,235,236].

**a. Particulate magnetoelectric composites**

The 0-3 connectivity type was initially used in the development of ceramic composites. In practice, the most commonly studied composite type involves mixing calcined powders of piezoelectric and magnetic phases, subsequently sintered under appropriate time and temperature conditions. These composites are very cost-effective due to their easy processing advantages. They also feature a considerable piezoelectric-magnetostrictive interfacial area, a crucial characteristic for a robust ME response. However, the inclusion of magnetostrictive components, primarily semiconductors or poor electrical insulators like ferrites, compromises the insulation of the composite leading to leakage issues. This makes the composite's ability to sustain the required fields for magnetization switching. In addition, a homogenous and well-dispersed magnetostrictive phase is required inside the piezoelectric matrix to prevent percolation and aggregation of the magnetostrictive nanoparticles. These occurrences have the potential to provide a conductive pathway for the charges produced inside the piezoelectric matrix. Consequently, the agglomeration limits the volume fraction of the magnetostrictive phase, thereby prevents from achieving the expected ME response. Additionally, porosity, cracks, and space-charge accumulation at interfaces are additional factors that make accurate

ME coupling measurement more difficult. All of these challenges combined make the fabrication of 0-3 magnetoelectric composites particularly challenging.

The first study in this area was conducted at the Philips Laboratory, where the quaternary system Fe-Co-Ti-Ba-O was developed using a unidirectional solidification process. In this composite, the observed magnetoelectric coupling was around 130 mV $cm^{-1}$ $Oe^{-1}$[237]. As the first investigation, the ME response is more than an order of magnitude larger than the maximum values found on single-phase compounds. In the early 2000s, interest in multiferroic materials has increased following the discovery of high magnetoelectric coupling in $Tb_{1-x}Dy_xFe_2$ (Terfenol-D) when combined with $PbZr_{1-x}Ti_xO_3$ (PZT) [238]. These initial investigations led to the exploration of a range of multiferroic composites. For example, Sagar et al. studied the magnetoelectric effect of $xCo_{0.9}Ni_{0.1}Fe_2O_4$–(1-x)[0.5($Ba_{0.7}Ca_{0.3}TiO_3$)-0.5($BaZr_{0.2}Ti_{0.8}O_3$)] synthesized by coprecipitation method, and obtained an important ME value (α=21.6 mV $cm^{-1}$ $Oe^{-1}$) in the composite containing 0.4CNF–0.4BCZT [91]. Likewise, Kumar et al. investigated the effect of CFO weight fraction on magnetoelectric properties of (1-x) $Ba_{0.85}Ca_{0.15}Zr_{0.1}Ti_{0.9}O_3$–$xCoFe_2O_4$ particulate composite. The ME coupling increases with ferrite fraction and the highest ME coupling of 14.8 mV $cm^{-1}$ $Oe^{-1}$ was observed for 0.6BCZT–0.4CFO composite [239]. The magnetoelectric effect in these types of composites is largely dependent on the sintering temperature. In that regards, Biman Kar et al. investigated this effect using $0.85Ba_{0.95}Ca_{0.05}Ti_{0.95}Sn_{0.05}O_3$–$0.15Ni_{0.7}Zn_{0.3}Fe_2O_4$ and observed that the coupling coefficient varies with sintering temperature. The sample sintered at 1300°C for 4 hours exhibited the highest ME value (α = 2.37 mV $cm^{-1}Oe^{-1}$) [240]. Other values of magnetoelectric coupling were reported in the literature (Table 3) like LSMO–0.5BCT-0.5BZT (α = 10.6 mV $cm^{-1}Oe^{-1}$) [241], BCTZ–CFO (α = 1.028 mV $cm^{-1}Oe^{-1}$) [242], $Ba_{0.9}Sr_{0.1}TiO_3$–$Ni_{0.9}Zn_{0.1}Fe_{1.98}O_{4-\delta}$ (BST–NZF) (α = 0.533 mV $cm^{-1}Oe^{-1}$) [243], and 0.8 (2.5BNT-22.5BKT-5BGT) –$0.2Ni_{0.7}Zn_{0.3}Fe_2O_4$ (α = 58.12 mV $cm^{-1}Oe^{-1}$) [244].

However, the theoretical coefficient for 0-3 particulate composites (2400 mV $cm^{-1}$ $Oe^{-1}$) is much higher than the experimental ones, mainly due to some defects. It refers to some specific problems related to the elaboration, including interdiffusion of phases, chemical reactions between the constituent starting materials during sintering processes yielding to the formation of undesirable phases (like $BaFe_{12}O_{19}$, $BaCo_6Ti_6O_{19}$, in the case of BT/CFO composites), and interfacial diffusion at atomic scale during the growing process [74,245]. In addition, high leakage currents are caused by the interdiffusion and/or chemical interactions between the magnetostrictive and piezoelectric phases during the sintering process. Moreover, the ceramics' electric poling becomes challenging and the ME interactions' strength decreases as a result of the leakage current issue [246]. On the other hand, randomly mixed magnetostrictive particles have a low percolation limit. To achieve a well-dispersed high concentration of this particles within composite ceramics, chemical methods, such as sol-gel and wet chemical processing, have recently been used to in situ synthesize homogeneously mixed piezoelectric and magnetic powders [84]. For instance, magnetoelectric particulate composites, xCFO–(1-x)BCZT (x= 30, 40, 50 wt%), were synthesized in-situ by a modified sol-gel method, and 0.4CFO–0.6BCZT showed enhanced magnetoelectric effect, exhibiting a maximum magnetoelectric coupling coefficient $\alpha_{ME}$=7.75 mV $cm^{-1}Oe^{-1}$ and $\alpha_{ME}$=161 mV $cm^{-1}Oe^{-1}$ at its resonance frequency [247]. In contrast, the CFO–BCZT particulate composite prepared by the mechanical mixing method showed a maximum $\alpha_{ME}$ of 118 mV $cm^{-1}Oe^{-1}$ at its resonance frequency. Notably, CFO–BCZT particulate composites prepared by the in-situ synthesis technique showed a 35% higher magnetoelectric coefficient than those produced via the mechanical mixing method [247]. The table 3 below summarizes recent findings on 0-3 magnetoelectric composites made of ferroelectric/ferrite involving different ferroelectric and ferromagnetic phases and various compositions, preparation and sintering techniques.

Magnetoelectric materials are becoming more and more common in biosensors and, more specifically, medical applications, though they require great biological compatibility and flexibility. The 0-3 polymer-based ME composites are particularly remarkable for meeting these requirements. Polyvinylidene fluoride (PVDF) and its derivatives, such as polyvinylidene fluoride-trifluoroethylene (P(VDF-TrFE)) and polyvinylidene fluoride-hexafluoropropylene (P(VDF-HFP)), are distinguished by their high flexibility, plasticity, and biocompatibility. Additionally, they exhibit good piezoelectric properties making them highly important in the field of magnetoelectricity. They are used as matrices to host various magnetostrictive particles. Recently, we developed new 0-3 composites in the form of nanofibers using the electrospinning method, with BCTSn as the piezoelectric phase and CFO as the magnetic phase. Piezoelectricity and magnetism were confirmed using PFM and M-H hysteresis loops, respectively. However, the magnetoelectric properties are still under investigation[248]. Chen et al., investigated the use of piezoelectric polymers in 0-3 ME composites in a recent overview listing some experimental studies from the bibliography [249], as shown in Figure 15a. The ME coefficient $\alpha_{E33}$ for these ME composites is found to be in the tens of mV $cm^{-1}$ $Oe^{-1}$ range. Significantly different ME coefficients are generated in the same component materials in different proportions [250,251], and different components in the same amount [252,253]. According to the law of composites, increasing one phase proportion generally influences the performance of the other phase. In this respect, the flexibility, adhesion, and piezoelectric properties of 0-3 polymer-based ME composites are affected by the composition of the piezoelectric matrix. A low percentage reduces these qualities, whereas an excessive amount diminishes the magnetic properties of the composites. As a result, component properties and proportions have a substantial impact on the ME properties of polymer-based composites (Fig. 15b). The primary challenge in enhancing the utilization of these composites lies in achieving a balance where excellent electrical and magnetic properties coexist. These materials are particularly well-suited for energy harvesting applications in compact and portable devices that require moderate magnetoelectric (ME) coupling and ease of fabrication. Additionally, in biomedical sensors, particulate composites can be effectively miniaturized for implantable applications where extremely high sensitivity is not a primary requirement. However, significant challenges remain, including achieving a uniform dispersion of a high concentration of the magnetic phase within the piezoelectric ceramic matrix, promoting favorable coherent interfaces, and ensuring sufficient bulk density while avoiding undesirable reactions and interfacial diffusion between the two ceramic phases. Further research is essential to address these issues and enhance the performance of these composites.

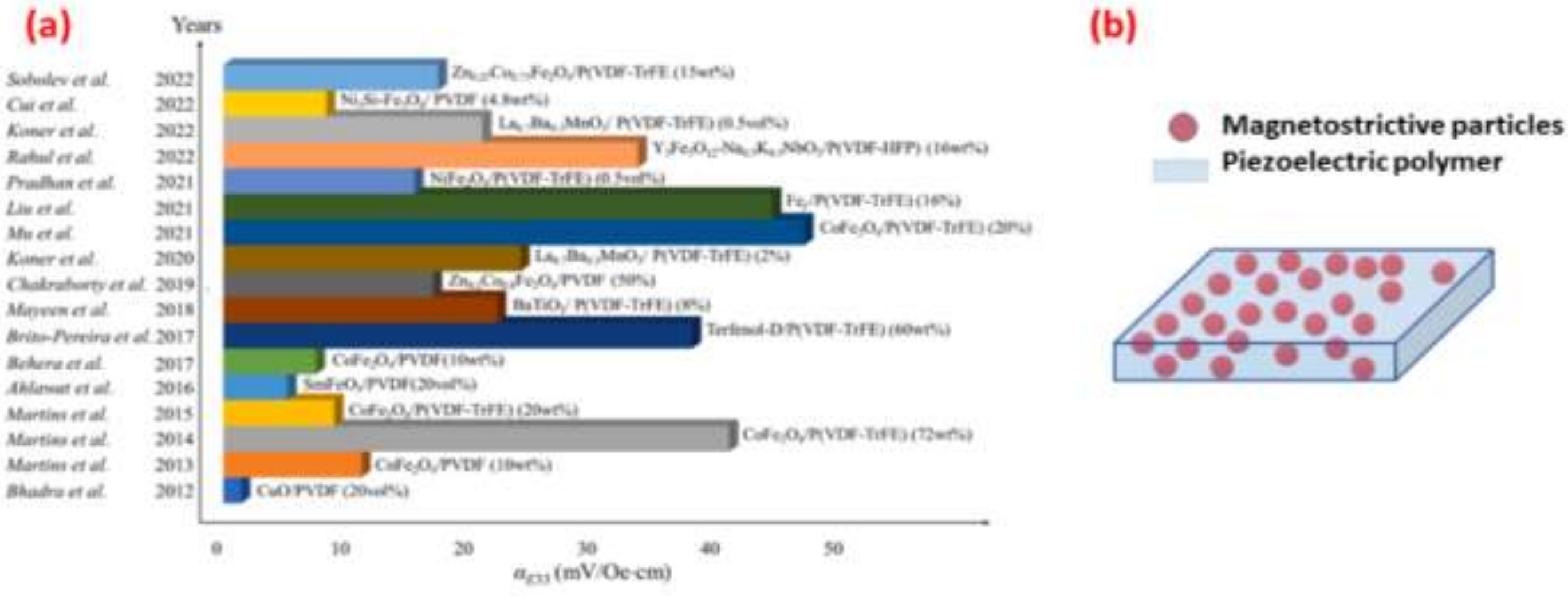

***Fig. 15.*** *(a) Research status of ME effect in 0-3 polymer matrix ME composites [249], (b)schematic of 0-3 based polymer ME composite.*

***Table 3:*** *Recent findings on 0-3 magnetoelectric composites.*

| Material | Synthesis method | Sintering temperature | Applied magnetic field DC (kOe) | Maximum ME response (mV $cm^{-1}$ $Oe^{-1}$) | References |
|---|---|---|---|---|---|
| $0.50BaFe_{12}O_{19}$–$0.40BaTiO_3$–$0.10CoFe_2O_4$ | Solid-state | 1200 °C /4 h | - | 4.56 | 254 |
| $0.95Ba_{0.99}Tb_{0.02}Ti_{0.99}O_3$–0.06 $Co_{0.94}Tb_{0.06}Fe_2O_4$ | Solid-state | 1000 ◦C /3 h | 1 | 3 | 255 |
| $0.88Ba_{0.95}Sn_{0.05}Ti_{0.95}Ga_{0.05}O_3$–$0.12CoFe_{1.8}Ga_{0.2}O_4$ | Solid-state | 1000 °C | 4 | 3.41 | 256 |
| $0.70Ba_{0.83}Ca_{0.10}Sr_{0.07}TiO_3$–$0.30MnFe_2O_4$ | Solid-state | 900°C / 1 h | 4 | 5.35 | 257 |
| 0.50 $BiFeO_3$–$0.50Pb(Zr_{0.53}Ti_{0.47})O_3$ | Solid-state | 680°C | - | 16.9 | 258 |
| $0.80Bi_{0.5}Na_{0.5}TiO_3$–$0.20Ni_{0.5}Zn_{0.5}Fe_2O_4$ | BNT : Sol-gel<br>NZFO: Auto-combustion | 1050 °C/ 1 h | 0.8 | 4.33 | 259 |
| $BaTi_{0.89}Sn_{0.11}O_3$–$CoFe_{1.9}Bi_{0.1}O_3$ | Sol-gel | 650 °C / 1 h | 1.4 | 7.8 | 260 |
| 0.40 $BaTiO_3$–0.60 $CoFe_2O_4$ ($Li_2CO_3$ 0.8 wt%) | Solid-state | 1100 °C /2 h | 3 | 35 | 261 |
| $0.85Ba_{0.95}Ca_{0.05}Ti_{0.95}Sn_{0.05}O_3$–$0.15Ni_{0.7}Zn_{0.3}Fe_2O_4$ | Solid-state | 1300 °C /4 h | 10 | 2.37 | 240 |
| $0.10CoFe_2O_4$–0.90[0.5Ba(Zr0.2$Ti_{0.8}$)$O_3$–-0.5($Ba_{0.7}Ca_{0.3}$)$TiO_3$] | BCZT: Sol-gel<br>CFO: Metallo-organic decomposition | 1150 °C /2 h | 8 | 1.02 | 242 |
| $0.60Pb_{0.4}La_{0.6}Zr_{0.6}Ti_{0.4}O_3$–$0.40CoFe_2O_4$ | PLZT: Solid-state<br>CFO: Sol-gel | 1160 °C /4 h | 1 | 0.45 | 262 |
| $0.85Pb(Zr_{0.53}Ti_{0.47})O_3$–$0.15(Ni_{0.5}Zn_{0.5})Fe_2O_4$ | PZT: modified Sol-gel<br>NZFO: Solid-state | 900 °C /4 h | 1 | 0.33 | 74 |
| $0.80BaZr_{0.25}Ti_{0.75}O_3$–$0.20Co_{0.9}Ni_{0.1}Fe_2O_4$ | Co-precipitation | 1100°C/30 min (microwave-sintering) | 4 | 2.71 | 263 |
| $0.55BaTiO_3$–$0.45CoFe_2O_4$ | Solid-state | 1200°C/24 h | 5 | 17 | 264 |
| $0.40Ba_{0.85}Ca_{0.15}Zr_{0.1}Ti_{0.9}O_3$–0.60 $Ni_{0.5}Zn_{0.5}Fe_2O_4$ | Sol-gel | 1250 °C /2 h | 3.5 | 2.55 | 265 |
| $Ba_{0.83}Ca_{0.10}Sr_{0.07}TiO_3$–$BiFeO_3$ | Solid-state | 700 °C /1 h | 15 | 3.61 | 266 |
| $Na_{0.5}Bi_{0.5}TiO_3$–$BaFe_{11}Co_{0.5}Ti_{0.5}O_{19}$ | PLZT: Solid-state<br>CFO: Sol-gel | 1100 °C /2 h | 0.4 | 59.81 | 267 |
| $(72.5Bi_{0.5}Na_{0.5}TiO_3$–$22.5Bi_{0.5}K_{0.5}TiO_3$–$5BiMg_{0.5}Ti_{0.5}O_3)$–$Ni_{0.7}Zn_{0.3}Fe_2O_4$ | Solid-state | 1050 °C /1 h | 0.8 | 58.12 | 244 |
| $0.90BiFeO_3$–$0.10GdFeO_3$ | Solid-state | 620 °C /7 h | 1 | 0.16 | 268 |
| $0.50(Ba_{0.85}Ca_{0.15})(Zr_{0.1}Ti_{0.9})O_3$–$0.50CoFe_2O_4$ | BCZT: Sol-gel<br>CFO: Auto combustion | 1300 °C /4 h | 2 | 6.85 | 269 |
| $0.70(Na_{0.41}K_{0.09}Bi_{0.5}TiO_3$–$Ba_{0.85}Ca_{0.15}Zr_{0.1}Ti_{0.9}O_3)$–$0.30(CoFe_2O_4)$ | Solid-state | 1100 °C /3 h | 2.3 | 3.54 | 270 |
| $0.80Na_{0.5}Bi_{0.5}TiO_3$–$0.20Ni_{0.5}Co_{0.5}Fe_2O_4$ | Solid-state | 1100 °C /3 h | 0.5 | 3.16 | 271 |
| $0.70Ba_{0.95}Ca_{0.05}Ti_{0.89}Sn_{0.11}O_3$–$0.30CoFe_2O_4$ | BCTSn: Sol-gel<br>CFO: Auto combustion | 1300 °C /4 h | 2.1 | 0.10 | 84 |
| $CoFe_2O_4$–$PbZr_{0.52}Ti_{0.48}O_3$ | Carbon templates | | | 3.39 | [272] |
| $0.60Ba_{0.85}Ca_{0.15}Zr_{0.1}Ti_{0.9}O_3$–$0.40CoFe_2O_4$ | BCZT: Solid-state<br>CFO: Sol-gel | 1200 °C | 2 | 14.8 | [239] |
| $0.85BaTiO_3$–$0.15Ni_{0.64}Zn_{0.36}Fe_2O_4$ | BT: Commercial powder<br>NZFO: Solide-state | 1000 °C /4 h | 0.8 | 2.99 | [273] |
| $0.75Ba_{0.95}Ca_{0.05}Ti_{0.95}Sn_{0.05}O_3$–0.25 $Ni_{0.7}Zn_{0.3}Fe_2O_4$ | Solid-state | 1300°C /4 h | | | [274] |
| x $NiFe_2O_4$ – (1-x) $BaTiO_3$ | Solid-state | 1400°C /2 h | | | [275] |
| $0.20CoFe_2O_4$–$0.80BaTiO_3$ | Co-precipitation | 600°C /8 h | 6 | 44.13 | [276] |
| $BiFeO_3$–$BaTiO_3$/$BaFe_{12}O_{19}$ | Solid-state<br>Cold sintering | 700°C /40 min | 3 | 0.39 | [277] |
| (x) $CoFe_2O_4$–(1−x) $Ba_{0.8}Sr_{0.2}TiO_3$ | Solid-state<br>Microwave sintering | 900°C /30 min | | | [278] |
| (1-x)$BaTiO_3$–x$CoFe_2O_4$ | Spark plasma sintering | 900 °C/ 12h | 1 | 0.0213 | [279] |
| 0.70PVDF–$0.30Ba_{0.7}Ca_{0.3}TiO_3$-$Co_{0.6}Zn_{0.4}Fe_2O_4$ | BCT/CZFO: Sol-gel | | 1.2 | 59 | [280] |
| $Mn_{0.5}Zn_{0.5}Fe_2O_4$–$PbZr_{0.5}Ti_{0.5}O_3$ (fluids) | MZFO: Solide-state<br>PZT: Sol-gel | | | 15.14 $10^3$ | [281] |
| $0.80Na_{0.41}K_{0.09}Bi_{0.5}TiO_3$-$Ba_{0.85}Ca_{0.15}Zr_{0.1}Ti_{0.9}O_3$–$0.20CoFe_2O_4$ | Solid-state | 1100°C/ 3 h | 2.3 | 3.58 | [270] |

### b. Laminated magnetoelectric composites

Compared to particulate composites, laminated magnetostrictive and piezoelectric composites have demonstrated much stronger magnetoelectric coupling. Praveen et al, investigated the magnetoelectric properties of BCZT–CFO composites with 0-3 and 2-2 connectivity types. The laminate composite exhibited a high magnetoelectric coefficient ($\alpha_{ME}$) of approximately 615 mV $cm^{-1}$ $Oe^{-1}$, that is six-fold greater than the particulate composite's $\alpha_{ME}$ of 104 mV $cm^{-1}$ $Oe^{-1}$ [282]. In bulk, the fabrication of these composites is generally achieved by bonding magnetostrictive and piezoelectric sintered pellets using silver adhesive epoxy followed by a heating process (approximately 150°C) to ensure proper adhesion between phases layers. This can also be accomplished by alternating multilayers of piezoelectric and magnetostrictive phases using the tape casting method [283]. Alternatively, it can be fabricated via cosintering of pressed powders of each phase, namely piezoelectric and magnetostrictive phases as shown in Figure 16a. This composite material can exist in various forms, such as bi-layers and tri-layers, where a piezoelectric phase is sandwiched between two magnetostrictive phases, as well as in multilayer configurations. Shara Sowmya et al. studied the effect of layer architecture on magnetoelectric effect in lead free laminates composite based on (0.5) BCT-(0.5) BZT and $NiFe_2O_4$. A maximum ME response of 980 and 1100 mV $cm^{-1}Oe^{-1}$ was obtained for the bi-layer BCZT/NFO and tri-layer NFO/BCZT/ NFO composites, respectively [284].

The horizontal alternating layers of piezoelectric and magnetostrictive phases results in avoiding the leakage problem observed in 0-3 particulate composites. Additionally, the separation between the insulating ferroelectric phase and the conducting magnetic phase leads to a higher degree of polarization and thereby exhibit better ME response and electric properties. However, it is important to note that the magnetostrictive-piezoelectric interfacial area in 2-2 composites is reduced compared to 0-3 composites, potentially leading to increased losses due to low-quality interface. Moreover, the different sintering temperatures of the phases in co-sintered laminate composites may also lead to poor densification and/or melting of one phase compared to the other, atom interdiffusion, and/or chemical reactions between two ceramic layers during the high-temperature sintering process. All of these issues provide a great challenge for academics and researchers trying to understand and overcome them.

As previously stated, 2-2 laminated composites may be produced using a variety of methods. The hot-pressing technique provides an alternate solution for avoiding inter-diffusion in co-sintered laminate composites. Inter-diffusion can change the magnetic and piezoelectric properties of phases, reducing the magnetoelectric response of composite ceramics. Ning Cai et al. reported high ME coupling as over 3000 mV $cm^{-1}$ $Oe^{-1}$ measured in the laminated composites with a Terfenol-D/PVDF composite layer sandwiched between two PZT/PVDF composite layers prepared via a simple hot-molding technique [285].

In some cases, laminated co-sintered composites have sintering issues due to sintering temperature differences between the piezoelectric and magnetostrictive phases, which prevents greater densification of one of the two phases (specifically, the phase with a high sintering temperature). Wang et al. elaborated magnetoelectric $Ba_{0.9}Ca_{0.1}Ti_{0.9}Zr_{0.1}O_3/CoFe_2O_4$ (BCZT/CFO) laminated composites (Fig. 16b). To overcome the sintering problem, a small amount of $Li_2CO_3$ (0.6 % wt) was added to the BCZT phase to lower its sintering temperature closer to that of CFO [286]. In addition, in a 2-2 laminate composite, the magnetoelectric effect depends on the measurement modes employed. There are four categorized types of modes (Fig. 16c): longitudinally magnetized and longitudinally poled (L–L) mode, longitudinally magnetized and transversely poled (L–T) mode, transversely magnetized and longitudinally poled (T–L) mode, and transversely magnetized and transversely poled (T–T) mode. In the case of a cofired ceramic composite, it is relatively easy to fabricate and pole the (L–T) and (T–T) modes. However, fabricating the (L–L) mode has proven significantly more challenging [5]. The longitudinal and transverse modes can exhibit distinct ME results within the same composite. For example, Srinivasan et al. measured the longitudinal (L-T) and (T-T) transverse (T-T) ME responses for PZT–$Ni_{1-x}Zn_xFe_2O_4$ (NZFO) (x = 0 - 0.5) laminated composites [287]. Figure 16d shows ME voltage coefficients versus H data for multilayer samples of NZFO–PZT. Even at low magnetic fields, the transverse coefficient is significantly higher than the longitudinal values [287].

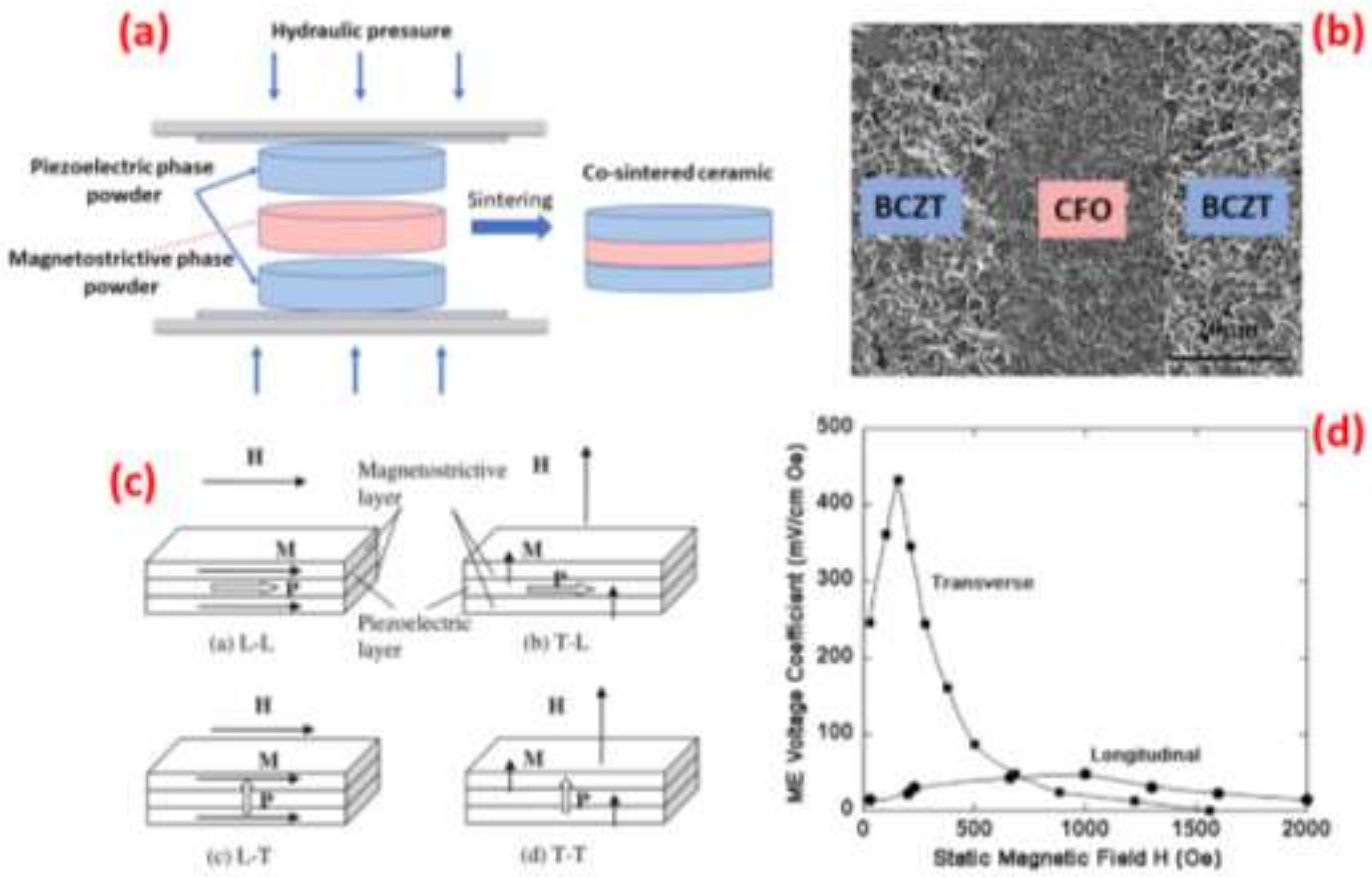

***Fig. 16.*** *(a) Schematic process for fabrication of co-sintered laminate composite. (b) SEM micrograph of 0.8BCZT/0.2CFO laminated composites (after [286]), (c) Schematic illustration of the various magnetoelectric coupling modes (reprinted with permission from [12]), (d) Transverse and longitudinal ME voltage coefficients versus H data for multilayer samples of NZFO–PZT(reprinted with permission from [288]).*

Thin films, bilayers, and multilayers are also classified as 2-2 type composites. These composites can be made chemically using the sol-gel spin coating process or physically with pulsed laser deposition, molecular beam epitaxy, and sputtering. Compared to other connectivities which rely on leakage current from the low resistance phase, 2-2 type connectivity has the benefit of being easier to fabricate and measure for direct ME coupling.

Moreover, the insulating ferroelectric layer in 2-2 type composites reduces short-circuits across the interface and allows for measurement of both direct and converse ME characteristics [289]. Very recently, a new 2-2 type composite was developed by Song et al. They utilized hexagonal boron nitride (h-BN) nanosheets dispersed in poly(vinylidene fluoride-trifluoroethylene) (P(VDF-TrFE)) to fabricate a thick piezoelectric film. This film is sandwiched between two layers of magnetostrictive material based on FeSiB alloy. The magnetoelectric voltage coefficient, generated by the designed ME composite with a width of 6 mm and a length of 22 mm, reached 57 V $cm^{-1}$. $Oe^{-1}$ [290]. Therefore, in this type of composite, several sensitive parameters must be taken into consideration to achieve strong coupling. For example, Kola et al. conducted magnetoelectric studies on $BaTi_{1-x}Sn_xO_3/NiFe_2O_4$ (x=0.07, 0.08, 0.09, and 0.10) lead-free bilayer laminated composites. The measured values of magnetoelectric coupling were 21.81, 25.37, 14.14, and 3.09 mV $cm^{-1}$ $Oe^{-1}$ for BTS7–NFO, BTS8–NFO, BTS9–NFO, and BTS10–NFO, respectively. In the same study, the effect of thickness is also explored. The ME measurements were conducted on BTS8/NFO laminate composites by varying the thickness of the BTS8 layer (0.5, 0.74, and 1 mm). The BTS8/NFO composites exhibit maximum values of $\alpha_{ME}$ at 25.7, 25.4, and 13.1 mV $cm^{-1}$ $Oe^{-1}$ for BST8 layer thicknesses of 0.55, 0.74, and 1 mm, respectively, around a 130 Oe field [77]. Similar to particulate composites, 2-2 composites are extensively used in energy harvesting applications, magnetic field sensing, and memory devices. For instance, Dai et al. designed a prototype vibration energy harvester incorporating a Terfenol-D/PZT/Terfenol-D laminate magnetoelectric transducer. This device demonstrated a power output of 1.05 mW across a 564.7 kΩ resistor under 1g acceleration at a resonant frequency of 51 Hz, showcasing the efficiency of 2-2 composites in converting mechanical vibrations into usable electrical energy [291]. Liu et al[292], investigated a single-phase $BiFe_{0.95}Co_{0.05}O_3$ films fabricated directly on PI substrates, demonstrating a magnetoelectric coupling of 0.0135 V $A^{-1}$, suggesting their great potential applications in flexible multiferroic memory devices. In addition, 2-2 composites are used in sensing applications due to their high sensitivity to magnetic fields, making them ideal for detecting weak magnetic signals. For example, a 2-2 FeCoSiB/AlN thin film demonstrated a detection limit of 1 pT/√Hz and a magnetoelectric (ME) coefficient as high as 6900 V $cm^{-1}$ $Oe^{-1}$ at the mechanical resonant frequency achieved in bending mode[293]. In addition, this type of composite has recently been utilized in the field of detection. For example, Victor V. Kuts et al. developed a 2D mapping technique for non-uniform magnetic fields (NMF) using magnetoelectric sensors. These ME sensors, based on $LiNbO_3$/Ni/Metglas, demonstrated a ME coefficient of 0.83 V $cm^{-1}Oe^{-1}$ without external biasing at 117 Hz[294]. However, they also present challenges that limit their effective use in their fields of application. For example, they face manufacturing challenges, such as the high-temperature co-firing process, which introduces issues like differential shrinkage rates and thermal expansion mismatches in laminated composites. Additionally, they exhibit relatively high resonance frequencies, which restrict their application in energy harvesting from industrial and household electrical devices.

The ME coupling coefficient of some 2-2 type composite systems is summarized in Table 4.

***Table 4:*** *Recent findings on 2-2 magnetoelectric composites*

| Materials | Processing route | Material Layer configurations | Applied magnetic field DC (kOe) | Maximum ME output ($\alpha_{E,31}$) (mV cm$^{-1}$ Oe$^{-1}$) | References |
|---|---|---|---|---|---|
| $CoFe_2O_4$ / $(Ba_{0.85}Ca_{0.15})$ $(Zr_{0.1}Ti_{0.9})O_3$ Thin film | Sol-gel spin coating | CFO/BCZT | 1 | 105 | 289 |
| $0.30(Ba_{0.85}Ca_{0.15})(Zr_{0.1}Ti_{0.9})O_3$/0.70 $La_{0.67}Ca_{0.33}MnO_3$ Laminated composite | Plasma activated sintering | BCZT/LCMO | 2 | 6.57 | 295 |
| $(Ba_{0.85}Ca_{0.15})$ $(Zr_{0.1}Ti_{0.9})O_3$ /$La_{0.67}Ca_{0.33}MnO_3$ Thin film | Pulsed laser deposition | BCZT/LCMO | 7 | 153.2 | 296 |
| 0.5BCT / 0.5BZT-$NiFe_2O_4$ Laminated composite | Bonding using silver epoxy | BCZT/NFO<br>NFO /BCZT/NFO | 0.18<br>0.52 | 980<br>1100 | 284 |
| $Bi_{3.4}La_{0.6}Ti_3O_{12}$ / $Ni_{0.7}Mn_{0.3}Fe_2O_4$ Laminated composite | Bonding using silver epoxy | BLT/NMFO<br>NMFO/BLT | 5.3<br>5.3 | 80<br>70 | 297 |
| $0.5BaTiO_3$ / $0.5CoFe_2O_4$ Laminated composite | Co-firing | BT/CFO/BT | 2.5 | 135 | 261 |
| $BaTi_{0.92}Sn_{0.08}O_3$ / $NiFe_2O_4$ Laminated composite | Bonding using silver epoxy | BST/NFO | 0.25 | 25.37 | 77 |
| $CoFe_2O_4$ / $(Ba_{0.85}Ca_{0.15})(Zr_{0.1}Ti_{0.9})O_3$ Laminated composite | Bonding using silver epoxy | CFO/BCZT/ CFO | 3 | 16 | 298 |
| $Co_{0.8}Ni_{0.2}Fe_2O_4$ / $K_{0.25}Na_{0.75}NbO_3$ Laminated composite | Bonding using silver epoxy | CNFO/KNN/ CNFO | 5 | 3.06 | 299 |
| $Pb_{0.895}Sr_{0.06}La_{0.03}(Zr_{0.56}Ti_{0.44})O_3$ / $Ni_{0.6}Zn_{0.4}Fe_2O_4$ Multilayer thick film | Tape casting | 4PSLZT/4 NZFO | self-biased | 230 | 72 |
| $Bi_{0.5}Na_{0.5}Ti_{0.98}Fe_{0.02}O_3$ / $NiFe_{1.98}Nd_{0.02}O_4$ Multilayer thin film | Sping coating | BNTF/NFNd | 2 | 169.7 | 76 |
| $Na_{0.4}K_{0.1}Bi_{0.5}TiO_3$ / $NiFe_2O_4$ | Bonding | NFO/NKBT/ | 0.4 | 80 | 300 |
| Laminated composites | using Silver epoxy | NFO | | | |
| $Ni_{0.8}Zn_{0.2}Fe_2O_4$ / $Ba_{0.85}Ca_{0.15}Zr_{0.1}Ti_{0.9}O_3$<br>Laminated composites | Bonding using silver epoxy | NZFO/BCZT/NZFO<br>BCZT/NZFO | 0.5<br>0.5 | 600<br>128 | 301 |
| $CoFe_2O_4$ / $(Ba_{0.85}Ca_{0.15})(Zr_{0.1}Ti_{0.9})O_3$<br>Laminated composite | Bonding using silver epoxy | CFO/BCZT<br>BCZT/CFO/BCZT | 2<br>2 | 68<br>43 | 302 |
| $0.5(Ba_{0.7}Ca_{0.3})TiO_3$-$0.5Ba(Zr_{0.2}Ti_{0.8})O_3$ /48%NiFe<br>Laminated composite | Bonding using silver epoxy | NF/BCZT | 0.2 | 67 | 303 |
| $CoFe_{1.7}Al_{0.3}O_4/Ba_{0.85}Ca_{0.15}Zr_{0.10}Ti_{0.90}O_3$<br>Laminated composite | Bonding using silver epoxy | CFAO/BCZT/CFAO<br>CFAO/BCZT | 0.25<br>0.25 | 100<br>48 | 304 |
| $72.5Bi_{0.5}Na_{0.5}TiO_3$–$22.5Bi_{0.5}K_{0.5}TiO_3$–$5BiMg_{0.5}Ti_{0.5}O_3$/ $Ni_{0.7}Zn_{0.3}Fe_2O_4$<br>Laminated composite | Thermal diffusion bonding | BNKMT/ NZFO<br>BNKMT/NZFO/BNKMT<br>NZFO/BNKMT/NZFO | 0.7<br>0.7<br>0.7 | 180.34<br>287.79<br>431.73 | 305 |
| $Pb_{(1-x-3y/2)}Sr_xLa_y(Zr_z,Ti_{(1-z)})O_3/Ni_{(1-x)}Zn_xFe_2O_4$<br>$Ba_{0.875}Ca_{0.125}(Zr_{0.125}Ti_{0.875})O_3$/ $Ni_{(1-x)}Zn_xFe_2O_4$<br>Laminated composite | Bonding using silver epoxy | PSLZT/NZFO<br>BCZT/NZFO | 0.3<br>2.5 | 22.4<br>8.9 | 306 |
| $BaTiO_3/NiFe_2O_4$<br>Laminated composite | Co-firing | BT/NFO/BT<br>NFO/BT/NFO | 2.5<br>2.7 | 10.5<br>8.43 | 307 |
| $La_{0.7}Sr_{0.3}MnO_3$/ $PbZr_{0.52}Ti_{0.48}O_3$<br>$La_{0.7}Sr_{0.3}MnO_3$/ $BaTiO_3$<br>Thin films | Pulsed laser deposition | LSMO/PZT<br>LSMO/BT | 1<br>1 | 63.75<br>57.75 | 308 |
| Metglas/PVDF | Theoretical study | | | 148.4 $10^3$ | 309 |
| $PbZr_{0.52}Ti_{0.48}O_3/Ni_{50.5}Mn_{27.9}Ga_{21.6}$/ $PbZr_{0.52}Ti_{0.48}O_3$ | Bridgeman method | PZT/FSMA/PZT | 7 | 215 | 310 |
| $Cu_{0.6}Co_{0.4}Fe_2O_4/PbZr_{0.58}Ti_{0.42}O_3$ | Screen printing method | CCFO/PZT/CCFO<br>PZT/CCFO/PZT | 0.2<br>0.4 | 91<br>83 | 311 |

### c. Core-shell magnetoelectric composites

Core-shell nanoparticles and 0-3 particulate composites are frequently misidentified in the literature. Multiferroic core-shell particles may be useful precursors for the production of 0-3 composites, which are made up of non-percolative inclusions of phase 1 that are evenly distributed inside a continuous matrix of phase 2, as shown in Figure 17.

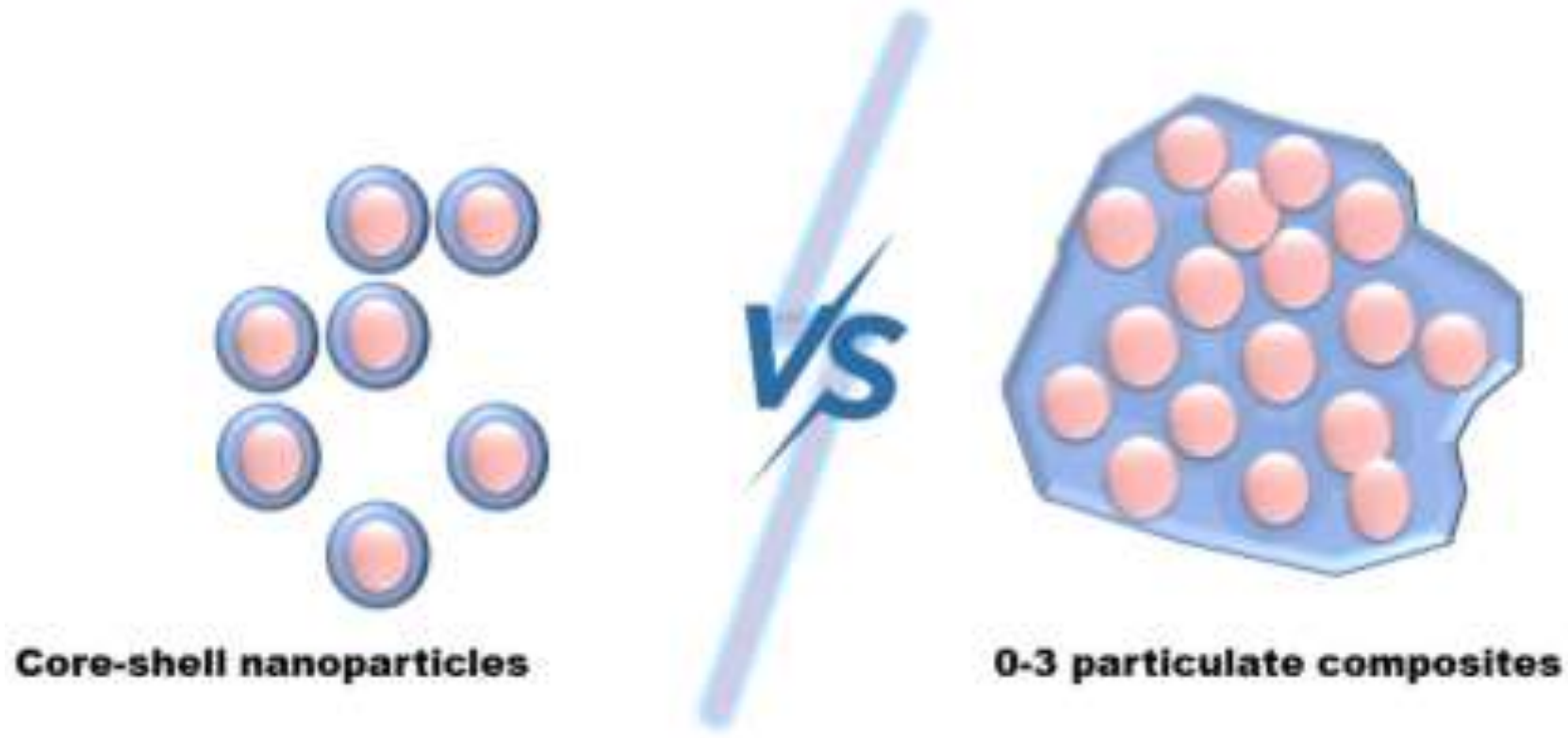

***Fig. 17.*** *Schematic diagram of core-shell nanoparticles and 0-3 particulate composites*

Multiferroic core-shell particles connectivity has a couple of potential advantages. First, the spherical shape, along with a well-adherent and dense shell, intimates the contact area between the two ferroic phases, enhancing strain-mediated coupling between polarization and magnetization. Theoretically, and if the crystal lattices of the two materials are in harmony, a strong coupling could be achieved by heteroepitaxial growth of the shell on the core surface, as in the case of ferrites' spinel structure and ferroelectric titanates' perovskite structure. In this regard, excellent coupling is demonstrated by thin-film heteroepitaxial structures made up of BT matrix-containing CFO pillars [312]. Additionally, compared to their bulk counterparts, core-shell nanostructures showed higher ME coupling. For example, CFO–BT core–shell nanostructures demonstrated a strain-mediated ME coupling coefficient of $\alpha_{ME}$ = 8.1 mV cm$^{-1}$ Oe$^{-1}$, which is 35 times more than that of bulk CFO–BT composite [313]. Magnetoelectric coupling is expected to be significantly greater in 1–3 composites and multiferroic coaxial nanofibers than in 0–3 composites and bilayered films (with 2–2 connectivity) of the same composition. However, making 1-3 composites is far more difficult than making 0-3 composites.

Second, by covering the conductive magnetic phase (core) with the insulating ferroelectric phase (shell), the dielectric losses can be lowered. Indeed, to achieve a high magnetoelectric coupling coefficient with beneficial magnetoelectric properties, a large resistance is required during the poling process. Three different types of multiferroic core–shell structures have been developed: spherical particles (Fig. 18a), coaxial fibers (Fig. 18b), and arrays of vertically aligned nanowires on a substrate (Fig. 18c).

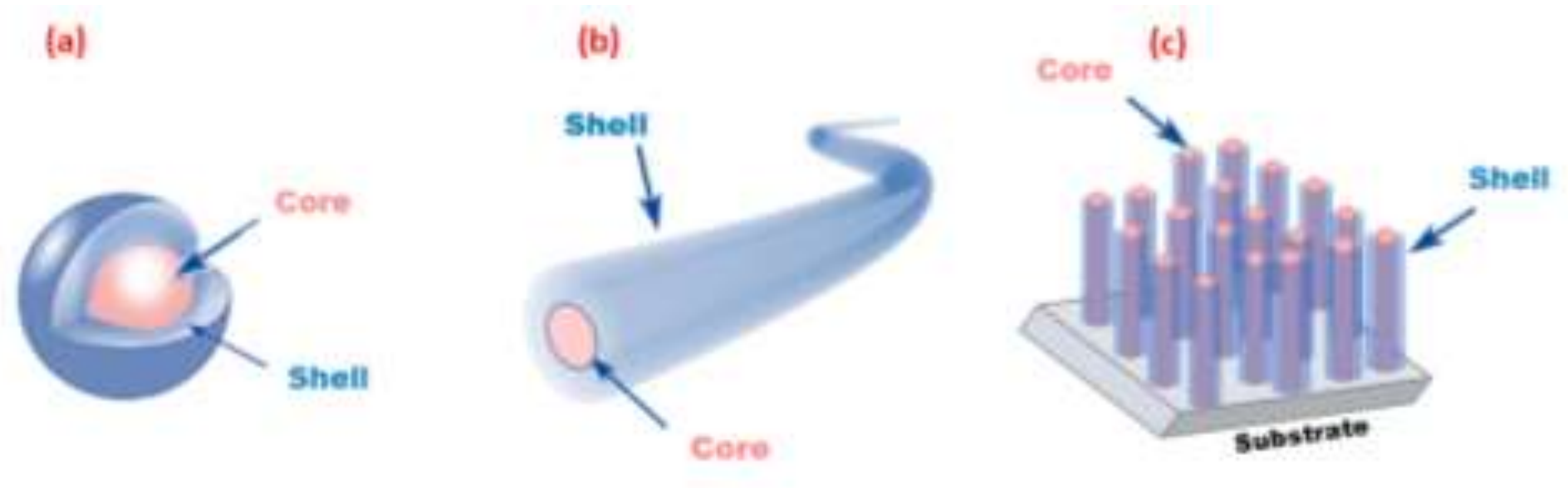

***Fig. 18.*** *Different types of multiferroic core–shell structures*

The most commonly used method for elaborating core-shell nanoparticles consists of dispersing magnetic/ferroelectric particles in a ferroelectric/magnetic solution. For instance, cobalt ferrite particles are dispersed in a sol obtained by dissolving a barium precursor (such as barium acetate, barium carbonate or $Ba(OH)_2$) and a titanium alkoxide (like n-butoxide or isopropoxide) in an aqueous solution of acetic acid to prepare $CoFe_2O_4$–$BaTiO_3$ core-shell composites [73,78]. This configuration is denoted as CFO@BT, where CFO represents the core and BT the shell (Fig. 19a). However, already mentioned, this method is the source of confusion between the 0-3 connectivity and Core-shell. This technique is called the elaboration of “in-situ” particulate composites. This technique was used by Monaji et al. to prepare [$xCoFe_2O_4$–$(1-x)(Ba_{0.85}Ca_{0.15})(Zr_{0.1}Ti_{0.9})O_3$] (x=30, 40, 50 wt%) magneto-electric particulate composites by mixing CFO particles into BCZT gel to avoid direct contact between the ferrite particles [247]. This reduces the problem of leakage current in situ and eliminates the low-conducting panel.

Perfect core-shell connectivity can be achieved through the hydrothermal method (Fig. 19b). This occurs as a result of the shell phase adhering to the core's surface during this reaction. Zhou et al., for instance, used a single-step hydrothermal treatment at 300°C to produce NFO@$BaTiO_3$ particles[314]. By hydrolyzing $TiCl_4$ and $Ba(NO_3)_2$ with NaOH solution, an amorphous precursor solution was produced in which the NFO particles (cores) were embedded. The pH of the mixture was then adjusted to 13. The formation of heteroepitaxial interfaces between $NiFe_2O_4$ and $BaTiO_3$ is necessary for the shell's development. The thickness of the shell may be adjusted by varying the quantity of ferrite particles. Another way for producing multiferroic core-shell particles is the self-assembly of independently produced particles. This method relies on the electrostatics and modified surface charge of each particle in each phase [315]. Mornet et al. have exploited the possibility of switching the surface charge of silica from positive to negative by functionalizing it with amino silane molecules to elaborate $BaTiO_3$-$\gamma$-$Fe_2O_3$ core-shell nanoparticles [316]. The as-prepared $BaTiO_3$ and $\gamma$-$Fe_2O_3$ particles were initially coated with a thin silica film. After functionalizing the $BaTiO_3@SiO_2$ particles, spontaneous formation of $(BaTiO_3@SiO_2)@(Fe_2O_3@SiO_2)$ assemblies occurs due to the interparticle electrostatic interaction when the two suspensions are mixed.

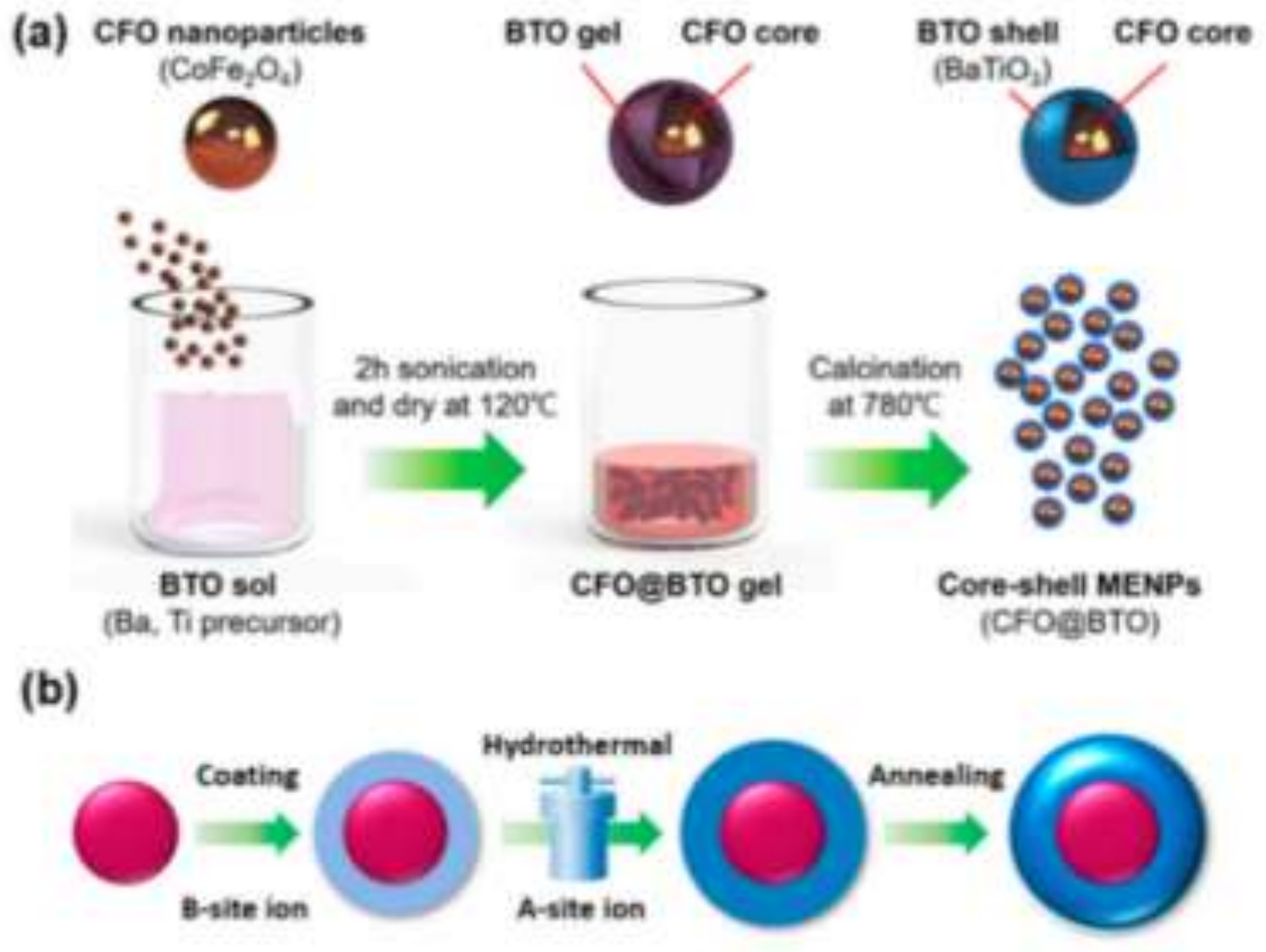

***Fig. 19.*** *Preparation of $CoFe_2O_4@BaTiO_3$ (CFO@BT) using the (a) sol–gel and (b) hydrothermal methods* [50].

The diameter and thickness of the composite nanoparticle's core and shell are key factors in this core-shell nanoparticle connectivity. A highly interesting theoretical investigation was undertaken utilizing finite element modeling (FEM) to simulate the experimental results conducted on CFO-BT based composites [317]. The study demonstrated how core diameter and shell thickness might influence the magnetoelectric coupling, as shown in Figure 20.

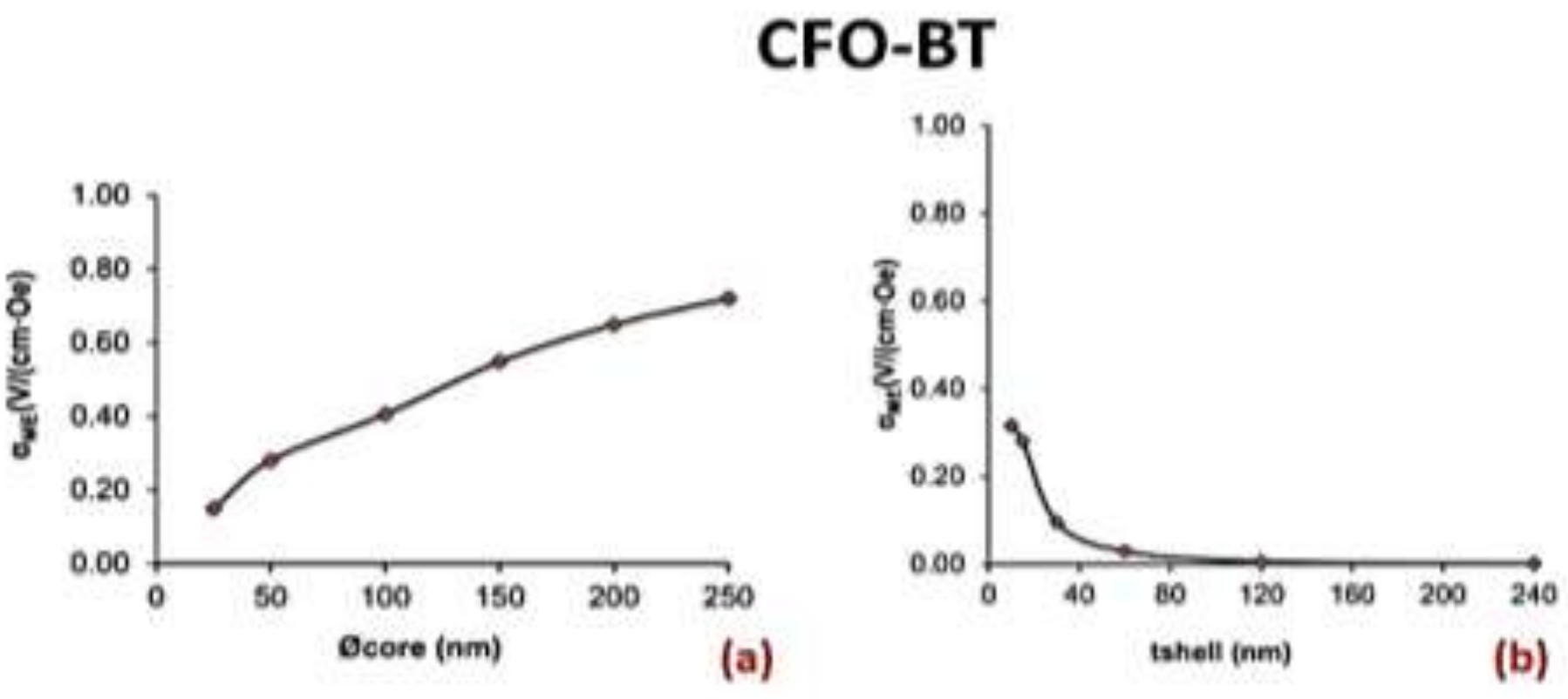

***Fig. 20.*** *Schematic representation of ME coupling dependence on (a) core diameter, (b) shell thickness, in CFO–BT core-shell nanoparticles* [317].

Electrospinning technique is a primary way to prepare composite multiferroic fibers. This technique is based on using a high-voltage electrostatic field to charge the surface of a precursor solution, inducing the ejection of a liquid jet through a spinneret. The prepared polymer solution is pumped into the tip of the needle, creating an electric field between the needle tip and the collector plate by applying high voltage in the system. When the surface tension in the liquid droplet is overcome by the force of the electric field, the droplet is distorted, forming the so-called "Taylor cone". This distortion results in an electrically charged jet that moves towards the collector, leading to the formation of thin fibers. The electrospun fibers can be tailored by adjusting various parameters related to the system including the distance between the needle and collector, voltage, and flow rate; as well as to the solution, such as concentration, conductivity, and viscosity. Environmental factors such as humidity and temperature are also important [318].

We have recently reported a study on $CoFe_2O_4$–$Ba_{0.95}Ca_{0.05}Ti_{0.89}Sn_{0.11}O_3$ core-shell nanofibers (CFO@BCTSn NFs) prepared by a sol-gel coaxial electrospinning technique (Fig. 21). The spinning process is detailed in the reference [319]. The BCTSn and CFO precursor sols containing PVP are separately injected in the inner (core) and outer (shell) parts of a coaxial spinneret assembled from handmade coaxial needle, as illustrated in the published paper (Fig.1). The coaxial fibers have a diameter of 150-250 nm and a well-defined core@shell geometry as seen in Fig. 2 and Fig. 3 of the published paper.

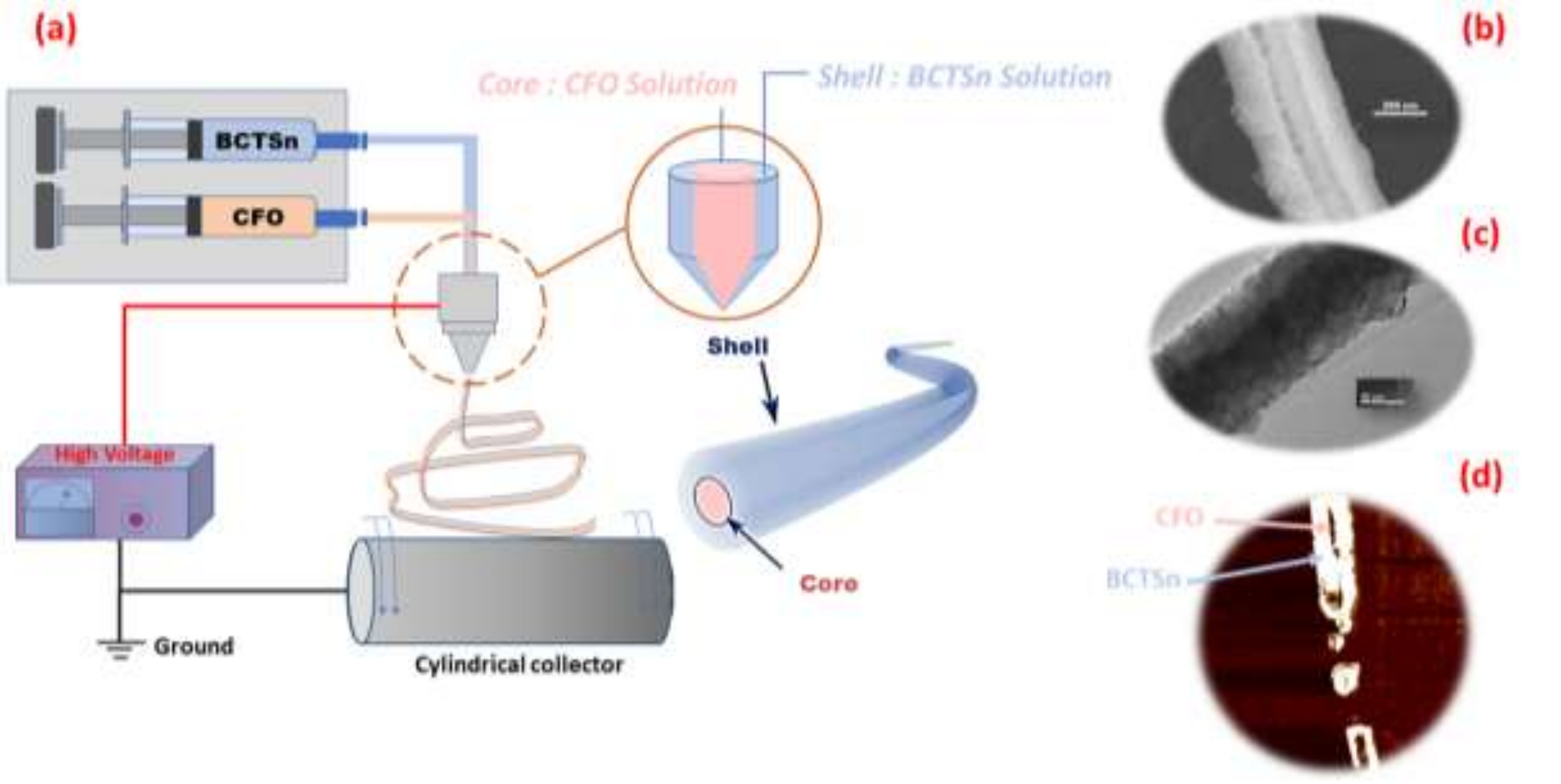

***Fig. 21.*** ***(a)*** *Coaxial electrospinning of core–shell nanofibers: setup schematics. (b), SEM image, (c) TEM image (inset, diffraction pattern acquired from the red-colored circle), and (d) lateral PFM of amplitude, of single CFO@BCTSn nanofiber (after* [319]*).*

Using templates or porous membranes as substrates is another approach for developing 1D core-shell structures involves. Indeed, the shell is prepared in the first step by depositing the solution into the vertical pores of the substrates, after which the core fills the void space inside the shell. For example, Ming et al. used porous Anodic Aluminium Oxide (AAO) templates to elaborate $NiFe_2O_4$–$Pb(Zr_{0.52}Ti_{0.48})O_3$ core-shell nanowire arrays [320]. The synthesis process consists of three steps: First, PZT nanotubes are elaborated by soaking AAO templates in the PZT gel precursors. Crystallization of PZT nanotubes is achieved by sintering at 650°C for 30 minutes. Second, the obtained PZT nanotubes were layered with copper by sputtering to make electrodes for the electrodeposition process of NiFe alloy. Finally, firing at 800°C for 24 hours gives $NiFe_2O_4$@PZT core-shell nanowires. The AAO template is removed using NaOH solutions. The process flow of producing ferrite core - PZT shell nanowire arrays using an AAO template is displayed schematically in Figure 22.

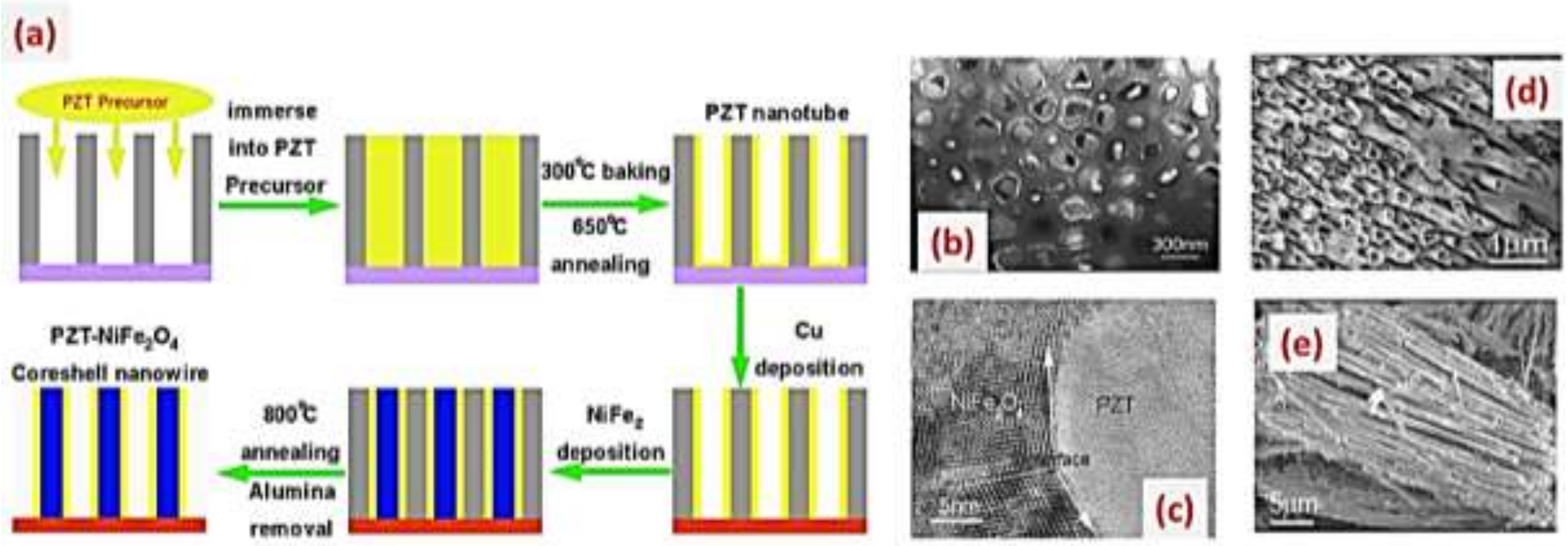

***Fig. 22.*** *(a) Schematic of the process flow for fabricating the ferrite core–PZT shell nanowire arrays, (b) Cross-section TEM image of core-shell PZT-NiFe2O4 nanowire in AAO template, (c) cross-section HRTEM image of core-shell PZT-*$NiFe_2O_4$ *interface area, (d) SEM images of open ends of PZT nanotubes with AAO template removed ,(e) core-shell PZT-*$NiFe_2O_4$ *nanowires released from AAO template* [320].

Cernea et al. used a polycarbonate membrane template to create 0.92BNT–0.08BT–CFO coaxial core-shell composite nanotubes [321]. First, BNT–BT nanotubes were elaborated by dropping sol precursor onto the surface of a 24 mm thick polycarbonate membrane template with a pore diameter of around 0.8 mm. Spin casting at 3000 rpm for 20 s was utilized to ensure that the sol was distributed uniformly on the membrane surface. The sol covers the channels' walls and forms tubes. After 15 minutes at room temperature, another quantity of sol was dropped on the membrane surface, and the process was repeated three times. Finally, the membrane with sol inside was exposed to air for 24 hours to complete transformation of the BNT–BT sol precursor in tubes gel. The same process was employed to prepare the $CoFe_2O_4$ tubes inside the obtained BNT–BT tubes. Thus, the $CoFe_2O_4$ sol precursor was dropped on the membrane surface, which contained BNT–BT gel tubes. After 24 hours in air at room temperature, the sol precursor of $CoFe_2O_4$ gels and gains strength inside BNT–BT gel tubes. The membrane was then dissolved in dichloromethane ($CH_2Cl_2$), and the coaxial BNT-BT/$CoFe_2O_4$ nanotubes were centrifuged out of the solution before being cleaned with isopropanol. BNT–BT/$CoFe_2O_4$ composite gel tubes were deposited on a Pt substrate and crystallized by annealing at 800 °C for 1 hour in oxygen. Figure 23 shows the production process scheme and SEM images for BNT–BT/$CoFe_2O_4$ core-shell nanotubes.

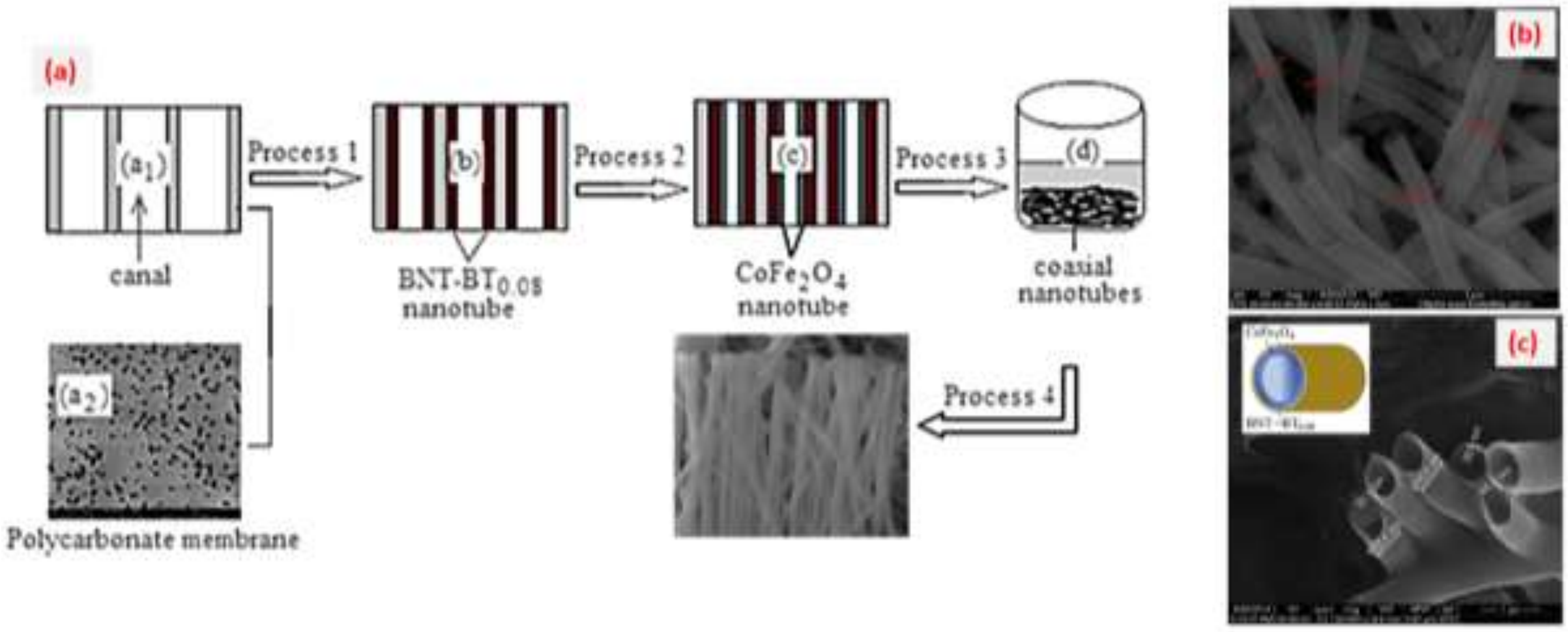

**Fig. 23.** *(a) Schematic of the fabrication process of BNT–BT/$CoFe_2O_4$ core-shell nanotubes, (b) and (c) SEM photomicrographs of the heterostructured BNT–BT/$CoFe_2O_4$ nanotubes, the inset of (c) is the schematic representation of the composite nanotube (reprinted with permission from [321]).*

In this type of connectivity, the magnetoelectricity induced by a magnetic field makes the core-shell structure useful for various biomedical, electrocatalytic applications and Internet of Things (IoT)-based devices. Core-shell structures are highly advantageous for biomedical applications due to their ability to intrinsically generate electric fields, which align with the natural electrical properties of living cells involved in recovery processes. A key feature is the wireless control of excitation via low-frequency magnetic fields. The unique piezoelectric shell enables on-demand drug release by leveraging the intrinsic electric field generated through the magnetoelectric effect. Additionally, magnetoelectric nanoparticles (MENPs) exhibit a strong affinity for various agents and drugs, making them ideal for targeted delivery, including antiviral treatments for HIV, CRISPR-Cas9/gRNA for latent HIV infection, and antiretroviral drug delivery[322]. In addition, core shell ME composites have promising applications in brain imaging, particularly in magnetic resonance imaging (MRI)[323]. MRI is commonly used for noninvasive brain examination and disease diagnosis, and core shell ME composites can enhance imaging quality by acting as contrast agents. In fact, by utilizing magnetoelectric coupling, the change in magnetization in core-shell composites detects voltage, which is then converted into a signal and further transformed into an image. This method is more sensitive because it not only produces a direct image based on data but can also quantify the changes. For instance, Nguyen et al.[37] utilized CFO@BTO nanoparticles for wireless stimulation of cortical neuron activity. Imaging techniques demonstrated that, under the influence of an external field, the MENPs could move freely within neurons and neural networks. Another example involves the use of a core-shell structure based on CFO@CTAB/PVDF, fabricated through electrospinning, to investigate its potential for treating chronic skin wounds, particularly diabetic wounds infected with *Staphylococcus aureus*[324]. Other biomedical applications benefit from magnetoelectric core-shell composites, such as brain stimulation and cell regeneration. For electrocatalysis applications, core-shell ME composites are used in water treatment, particularly for degrading organic pollutants in wastewater. Unlike traditional photocatalysts and magnetic nanostructures, core-shell ME composites can induce redox reactions to generate hydroxyl and superoxide radicals, significantly enhancing the degradation efficiency of water pollutants-up to 97% within an hour. Mushtaq et al.[325] investigated the electrochemical processes induced by CFO@BFO nanoparticles under an alternating magnetic field. By comparing the catalytic degradation curves of Rhodamine B (RhB) for pure CFO, pure BFO, and core–shell structures under an external field, they observed that the magnetoelectric effect in the core–shell structure achieved a remarkable 97% removal rate of RhB. MENPs have also demonstrated the ability to remove common pharmaceuticals from water with up to 85% efficiency. This high efficiency makes MENPs promising catalysts for wastewater purification[325]. Bitna Bae et al, obtained an optimized ME composite film based on CFO-BT core shell, and achieves an ME coefficient of 684 mV $cm^{-1}Oe^{-1}$ at a frequency near 600 Hz suggesting that the fully flexible ME generator could serve as a permanent power source for IoT-based magnetic field sensors[326].

Like other connectivities, core-shell structures also face challenges, including the size-dependent decline in piezoelectric properties. Reducing the piezoelectric phase to the nanoscale diminishes the asymmetric lattice ratio and increases reliance on dielectric properties, making improved material design essential for achieving a strong magnetoelectric effect. Additionally, nanoparticle aggregation during synthesis must be avoided, as it affects size uniformity and compromises the clarity of core-shell structures, thereby impacting the reproducibility of ME measurements. Current synthesis methods, such as sol–gel and hydrothermal techniques, are not scalable for industrial mass production, emphasizing the need for more efficient and reproducible approaches. Accurate quantification of the ME effect also requires refined measurement techniques. Furthermore, as core-shell structures are increasingly applied in biomedical fields, long-term in vitro and in vivo studies are critical to enhancing the understanding of their working mechanisms and ensuring biocompatibility. Measuring magnetoelectric coupling in these composites remains a significant challenge, leading most studies to focus on qualitative confirmation of this coupling rather than in-depth characterization. As a result, discussions in the literature often prioritize synthesis procedures over comprehensive analysis of the resulting composites.

Magnetoelectric materials (especially NPs core shell) offer immense potential for diverse future applications, particularly in nanomedicine and nanobiotechnology. These include wireless brain stimulation, real-time mapping and recording of neural activity, targeted delivery across the blood–brain barrier, tissue regeneration, high-specificity cancer treatments, and rapid molecular-level diagnostics. For instance, Singer et al. demonstrate that alternating magnetic fields can power millimeter sized ME stimulators in freely moving rodents. The extreme miniaturization made possible by this technology lays the foundation for a new class of minimally invasive bioelectronics [36].

Their unique coupling of magnetic and electric properties enables precise nanoscale control, making them ideal for integration into advanced biomedical devices. For example, magnetoelectric nanoparticles can facilitate non-invasive therapies by remotely triggering electric signals via magnetic fields, minimizing the need for surgical interventions. Their use in biosensors enhances sensitivity and specificity for biomarker detection, enabling early disease diagnosis.

Beyond healthcare, magnetoelectric materials are being investigated for energy-efficient memory devices, spintronics, and environmental monitoring, underscoring their transformative potential across scientific and technological domains. They also hold promise for breakthroughs in areas such as quantum computing, advanced robotics, and sustainable energy solutions. However, these niche applications fall outside the scope of this review.

***Table 5:*** *ME coupling of some core-shell nanofibers and nanoparticles composites.*

| Materials | Processing route | Composite morphology | Applied magnetic field DC (kOe) | Maximum ME output (mV $cm^{-1}$ $Oe^{-1}$) | References |
|---|---|---|---|---|---|
| $CoFe_2O_4$ @ $BaTiO_3$ | Co-precipitation method | Core–shell nanoparticle | 1 | 3.4 | [73] |
| $CoFe_2O_4$ @ $Pb(Zr_{0.2}Ti_{0.8})O_3$ | Sol-gel electrospinning | Coaxial nanofibers | - | 1.2 × $10^{-8}$ s/m | [327] |
| $CoFe_2O_4$ @ $Ba_{0.95}Ca_{0.05}Ti_{0.89}Sn_{0.11}O_3$ | Sol-gel electrospinning | Coaxial nanofibers | 10 | 346.4 | [319] |
| $CoFe_2O_4$ @ $BaTiO_3$ | Sol-gel electrospinning | Coaxial nanofibers | - | 1.2 × $10^4$ | [328] |
| $CoFe_2O_4$ @ $(Ba_{0.7}Ca_{0.3})(Zr_{0.2}Ti_{0.8})O_3$ | Sol-gel electrospinning | Coaxial nanofibers | - | - | [329] |
| $CoFe_2O_4$ @ $Pb(Zr_{0.52}Ti_{0.48})O_3$ | Sol-gel electrospinning | Coaxial nanofibers | - | - | [330] |
| CFO @ BT @ PDA/P(VDF-TrFE) | Sol-gel | Composite films | | 150.58 | [331] |
| $CoFe_2O_4$ @ $BaTiO_3$ | Hydrothermal reaction and polymer-assisted deposition | Self-assembled feather-like | 3.1 | 51.8 | [332] |
| $NiFe_2O_4$ @ $PbZr_{0.52}Ti_{0.48}O_3$ | Sol-gel electrospinning | Coaxial nanofibers | - | - | [333] |
| $NiFe_2O_4$ @ $BaTiO_3$ | Sol-gel electrospinning | Coaxial nanofibers | - | 0.4 | [96] |
| $La_{0.7}Sr_{0.3}MnO_3$ @ $BaTiO_3$ | Sol-gel method | Core–shell nanoparticle | 1 | 54.5 | [334] |
| $SrFe_{12}O_{19}$ @ $PbZr_{0.52}Ti_{0.48}O_3$ | Sol-gel electrospinning | Coaxial nanofibers | 7.2 | 3.227 | [335] |
| $SrFe_{12}O_{19}$ @ $BaTiO_3$ | Sol-gel electrospinning | Coaxial nanofibers | 7.1 | 3.852 | [335] |
| BNT–BT0.08 @ $CoFe_2O_4$ | Sol-gel-polycarbonate membrane template | Core-shell composite nanotubes | - | - | [321] |
| $CoFe_2O_4$ @ $Pb(Zr_{0.52}Ti_{0.48})O_3$ | Sol-gel electrospinning | Coaxial nanofibers | - | 2.95 $10^4$ | [336] |
| $NdFeO_3$ @ $PbZr_{0.52}Ti_{0.48}O_3$ | Sol-gel electrospinning | Coaxial nanofibers | - | - | [337] |
| $CoFe_2O_4$ @ $BaTiO_3$ | Wet chemical route | Core–shell nanoparticle | 10 | 180 | [338] |
| $Ba_2Zn_2Fe_{12}O_{22}$ @ $PbZr_{0.52}Ti_{0.48}O_3$ | Sol-gel electrospinning | Coaxial nanofibers | - | - | [339] |
| $Ba_{0.85}Sr_{0.15}TiO_3$ @ $Ni_{0.75}Zn_{0.25}Fe_2O_4$ | Microwave-assisted hydrothermal technique | Core–shell nanoparticle | 3.7 | 12.41 | [340] |
| $CoFe_2O_4$ @ $BaTiO_3$ | Sol-gel (Cores are prepared by hydrothermal method) | Core–shell with CFO core: Spherical Rectangular Nanowire | | $5.50 \times 10^3$<br>$23.93 \times 10^3$<br>$14 \times 10^3$ | [75] |
| $CoYb_{0.1}Fe_{1.9}O_4$ @ $BaTiO_3$ | Sol-gel (Core is prepared by Co-precipitation method) | Core–shell nanoparticle | - | - | [341] |
| $0.12Ni_{0.5}Co_{0.5}Fe_2O_4$ @ $0.88BaTiO_3$ | Sol-gel (Cores are prepared by hydrothermal method) | Core–shell nanoparticle | 1.1 | 20 | [342] |
| $CoFe_2O_4$-$BaTiO_3$/PVDF | Sol-gel (Cores are prepared by co-precipitation method) | Core–shell nanoparticle embedded in the PVDF matrix | 0.75 | 684 | [326] |
| $Ba_2Zn_2Fe_{12}O_{22}$–$PbZr_{0.52}Ti_{0.48}O_3$ | Sol-gel electrospinning | Coaxial nanofibers | | | [339] |

The statement concludes that while the connectivity of composite systems is crucial in determining their magnetoelectric properties, it cannot definitively claim that the 2-2 connectivity is superior to 0-3 or other configurations. Various parameters contribute to achieving strong coupling, some of which are challenging to control.

For example, in a composite of $CoFe_2O_4$ and $BaTiO_3$, the magnetoelectric coupling values differ across different connectivity types. In particulate composites, due to the difficulty of polarization of samples caused by the high conductivity of cobalt ferrite, the achieved values of magnetoelectric coefficients were in the range of 0.04-56 mV $cm^{-1}$. $Oe^{-1}$ (Table 3). Except for one unique case, Lirong Wang measured an extremely strong magnetoelectric coupling in the BT-CFO composite, approximately 587.3 mV $cm^{-1}$. $Oe^{-1}$ at 1.8 kHz, which may not be the only example of such composites [343].

As already stated, the low ME coupling in these composites is owing to the CFO phase distribution in the BT matrix, which is typically not uniform. Because of its high conductivity, the CFO phase creates percolated clusters that make it difficult to pole these samples.

To improve contact between the two phases, increase insulating characteristics, and enlarge the interface area, the concept of a core-shell structure was proposed, in which particles of CFO are fully separated by BT shells. However, the core-shell structure proved challenging to maintain during higher-temperature ceramic sintering. For instance, the core-shell nanoparticles composite of 0.5BT–0.5CFO had a maximum ME coupling of 1.5 mV $cm^{-1}Oe^{-1}$, indicating comparatively low values[344]. Significant values of ME coupling, such as 5.50 $10^3$ mV $cm^{-1}Oe^{-1}$ [75] and 180 mV $cm^{-1}Oe^{-1}$ [338], are extremely uncommon, as previously indicated.

Synthesis of composites with core-shell structure was also reported by Duong et al., and the ME coupling value reported to be 3.53 mV $cm^{-1}Oe^{-1}$. However, the authors did not provide evidence of the core-shell structure for either the powders or the ceramics [73]. On the other hand, CFO–BT core-shell structures were successfully synthesized by Betal et al. However, the ME coupling values were not measured for these nanopowders [345]. It is worth noting that these measured values are still lower than the theoretical values for different compositions of BT and CFO predicted by Fiocchi et al [317].

In (2-2) laminated composites, where ferromagnetic layers directly contact ferroelectric layers, and the ferromagnetic layers are completely insulated from each other, the value of ME coupling measured for co-sintered CFO–BT composites is 135 mV $cm^{-1}Oe^{-1}$ [261], which is larger than that for particulate composites but still far from the theoretical value of 800 mV $cm^{-1}Oe^{-1}$ for (2-2) composites [346]. For coaxial nanofibers of CFO–BT may exhibit a ME effect value in the range of 0.1 to 1 V $cm^{-1}Oe^{-1}$, which is larger than the theoretical value for (2-2) connectivity composites of CFO–BT.

This comparison suggests higher coupling in coaxial nanofibers composites. However, changing the magnetic phase to NFO can significantly impact the coupling. For instance, a study showed a magnetoelectric coupling of 0.4 mV $cm^{-1}$ $Oe^{-1}$ in NFO/BT core-shell fibers, emphasizing the complexity of magnetoelectric coupling and the need to control all influencing parameters for a comprehensive understanding.

Subsequently, the primary challenges here are the synthesis methods that are influenced by many physicochemical parameters. Other problems include measuring methods and parameters. For most systems, these factors have yet to be thoroughly defined, such as the relation between the converse and direct ME effect measurements, the frequency dependence of the ME coupling, and the interrelationship between the macroscopic materials properties and the magnetoelectric effect.

## V Summary and outlook

Magnetoelectric materials are a huge source of technological applications because they can simultaneously display tunable mechanical, magnetic, and electric, properties. Nevertheless, there remain a lot of challenges and limitations that require further research. Several studies have highlighted the need to better understand the fundamental mechanisms underlying multiferroic behavior. The coupling between a material's electrical and magnetic properties is complicated and still poorly understood, making it difficult to predict and design materials for specific applications. More research is required to develop new techniques for controlling and modifying the properties of these materials. Furthermore, the nonlinear behavior of multiferroic materials makes controlling their properties more complicated. Another significant challenge is the difficulty in synthesizing and characterizing ME materials. In fact, ME materials frequently have complex crystalline structures that need the employment of highly developed synthesis process. This can result in high fabrication costs and limited scalability. As a result, new synthesis techniques and less complicated treatment methods with low costs are required. Furthermore, ME materials must be able to be fabricated at micro and nanoscales without losing functionality.

The investigations on ME composites described until now reveal promising outcomes with significant development, along with the demonstration of several prototype devices; however, numerous remaining issues need to be addressed before genuine device applications.

One of the most significant aspects to consider when developing magnetoelectric composites is the selection of materials. These materials must possess appropriate mechanical, thermal, and chemical properties, as well as strong magnetoelectric coupling. Indeed, prioritizing the discovery of new room temperature magnetoelectric materials based on theoretical predictions, exhibiting strong cross-coupling between magnetization and polarization, low leakage current, and high remanent magnetization is essential. Furthermore, it is necessary to investigate new FE and magnetic domain structures, such as magnetic skyrmions and polar vortices, as they may lead to various types of magnetoelectric coupling. Elucidating the effect of microstructure on magnetoelectric coupling is of utmost importance. In addition, understanding how the nano/microstructure affects magnetoelectric coupling is critical.

Therefore, advanced characterization tools are required to study multiferroic and magnetoelectric properties at the nanometer and atomic length scales. Finally, it is critical to ensure successful on-chip integration, compatibility with various process technologies, and performance optimization of ME devices. All of these features are required for ME materials to be effectively used in actual device applications.

In conclusion, although current multiferroic materials face challenges and limitations, ongoing research aims to create novel materials with enhanced properties. In the coming years, the continuous advancements in synthesis, processing, and characterization techniques, ME materials are poised to offer exciting area of research and tremendous potential for diverse applications.

## Acknowledgments:

The authors gratefully acknowledge the generous financial support of HORIZON-MSCA-2022-SE H-GREEN (No. 101130520), MSCA-2020-RISE-MELON (No. 872631), and Slovenian Research Agency (research projects N2-0212, J2-3058 and research core funding P2-0105).